\documentclass[journal ]{new-aiaa}
\usepackage{amsmath}
\usepackage[utf8]{inputenc}
\usepackage{textcomp}
\usepackage{graphicx}
\usepackage[version=4]{mhchem}
\usepackage{siunitx}
\usepackage{longtable,tabularx}
\usepackage{hyperref}
\usepackage{latexsym}
\usepackage{multirow}
\usepackage{makecell}
\usepackage{graphicx}
\usepackage{float}
\usepackage{subfigure}
\usepackage{subcaption}

\usepackage{algorithm}
\usepackage{algpseudocode}
\title{Robust Angles-Only Initial Relative Orbit Determination Using Polynomial Optimization}

\author{Xingyu Zhou\footnote{Ph.D. Candidate, School of Aerospace Engineering; \url{zhouxingyu@bit.edu.cn}.}}
\affil{Beijing Institute of Technology, 100081 Beijing, People’s Republic of China}

\author{Malcolm Macdonald\footnote{Professor, Department of Electronic and Electrical Engineering; \url{malcolm.macdonald.102@strath.ac.uk} (Associate Fellow AIAA).}}
\affil{University of Strathclyde, Glasgow, Scotland G1 1XJ, United Kingdom}

\author{Roberto Armellin\footnote{Professor, Te Pūnaha Ātea - Space Institute, 20 Symonds Street, Auckland Central; \url{roberto.armellin@auckland.ac.nz} (Member AIAA).}}
\affil{University of Auckland, Auckland 1010, New Zealand}

\author{Dong Qiao\footnote{Professor, School of Aerospace Engineering; \url{qiaodong@bit.edu.cn} (Corresponding Author).} and Xiangyu Li\footnote{Associate Professor, School of Aerospace Engineering; \url{lixiangy@bit.edu.cn}.}}
\affil{Beijing Institute of Technology, 100081 Beijing, People’s Republic of China}

\begin{document}

\maketitle

% \begin{abstract}
% These instructions give you guidelines for preparing papers for AIAA Technical Journals using \LaTeX{}. If you previously prepared an AIAA Conference Paper using the Meetings Papers Template, you may submit using the Meetings Papers Template so long as the text is double-spaced.  Carefully follow the journal paper submission process in Sec.~II of this document. Keep in mind that the electronic file you submit will be formatted further at AIAA. This first paragraph is formatted in the abstract style. Abstracts are required for regular, full-length papers and express articles. Be sure to define all symbols used in the abstract, and do not cite references in this section. The footnote on the first page should list the Job Title and AIAA Member Grade (if applicable) for each author.
% \end{abstract}

%%%%%%%%%%%%%%%%%%%%%%%%%%%%%%%%%%%%%%%%%%%%%%%%%%%%%%%%%%%%%%%%%%%%%%%%%%%%%%%%%%%%%
%%%%%%%%%%%%%%%%%%%%%%%%%%%%%%%%%%%%%%%%%%%%%%%%%%%%%%%%%%%%%%%%%%%%%%%%%%%%%%%%%%%%%
\section{Introduction} \label{sec1}
\lettrine{R}{elative} orbit determination (ROD) is a key capability for space surveillance, autonomous proximity operations, rendezvous, and non-cooperative servicing missions \cite{Song2022AA,Sun2022JGCD}. It is commonly divided into two stages: initial relative orbit determination (IROD) \cite{LeGrand2015}, which provides an initial estimate of the relative state over a finite observation arc, and orbit refinement, which sequentially improves this estimate as new measurements become available \cite{Zhou2025IEEE}. Among the available measurement types, angle-only observations are particularly attractive because passive optical sensors provide low-cost line-of-sight (LOS) measurements without requiring target cooperation \cite{Liu2024IEEE, Zhou2023AA}. However, angles-only IROD suffers from a fundamental range ambiguity: under linearized dynamics, a body-centered sensor, and no maneuvers, the relative distance is unobservable \cite{Geller2014}. Existing methods mitigate this difficulty mainly through four routes: exploiting richer dynamical information, such as nonlinear motion or perturbations \cite{Ardaens2019}; introducing maneuvers as excitation inputs \cite{Wijayatunga2025}; creating geometric parallax through sensor offsets, attitude motion, or multi-sensor/multi-platform configurations \cite{Geller2014}; and augmenting LOS measurements with additional observables, such as apparent target size or other visual features \cite{Takahashi2021, Zhou2025JGCD}. Among these, nonlinear-dynamics-based methods are attractive because they enhance observability without additional resources \cite{Willis2024}.

Several nonlinear-dynamics-based approaches have been developed, including second-order models in Cartesian coordinates \cite{Willis2024}, curvilinear formulations in cylindrical or spherical frames \cite{Geller2017}, methods based on relative orbital elements \cite{Gaias2014, Sullivan2021}, and data-driven mappings from LOS histories to relative states \cite{Gong2023AA}. 
A particularly general and efficient framework is the overdetermined eigenvector approach (OEA) \cite{Kulik2024}. It constructs a cross-product-based objective from LOS observations and approximates the relative motion using second-order state transition tensors (STTs). The relative-state direction is first estimated using the linear term of the STTs, and the scale is then recovered by incorporating second-order terms. This enables angles-only IROD under arbitrary nonlinear dynamics without requiring maneuvers, sensor offsets, or prior range information. Nevertheless, its accuracy and robustness may degrade when the measurement noise or relative distance is large. 
This limitation mainly stems from three factors. First, the second-order STT approximation is itself limited in accuracy, while the direction estimate still relies on the linear term. 
Second, the inherent structure of the cross-product-based objective function makes the zero solution a local optimum, which may attract the optimization under unfavorable conditions.
Third, the OEA does not explicitly incorporate measurement weighting. A related weighted strategy has been explored for orbit refinement, where noticeable accuracy improvement was observed, but LOS measurements intrinsically contain only two degrees of freedom (DoF), making the resulting weighting matrix prone to numerical singularity and poor robustness under large-noise conditions. \cite{Sinclair2020}.

To this end, this Note develops an accurate and robust angles-only IROD method applicable to arbitrary nonlinear dynamics. The main contributions are threefold. 
First, the relative motion is approximated by high-order Taylor polynomials, thereby improving the approximation accuracy. Based on the LOS cross-product residual, the angles-only IROD problem is formulated as a polynomial optimization problem, which can then be solved efficiently using a recursive polynomial optimization (RPO) procedure. 
Second, a reduced-order weighting strategy is developed to avoid the numerical singularity induced by the limited DoF of LOS measurements. 
Third, a zero-solution-avoidance constraint is introduced together with an adaptive threshold selection mechanism, enabling robust suppression of the undesired zero solution under different observation geometries and noise levels. 
The proposed method is assessed against several representative approaches in terms of IROD accuracy, robustness, and computational cost.

The remainder of this paper is organized as follows. Section~\ref{sec2} introduces the preliminaries. The IROD methodology, together with the three contributions, are detailed in Sec.~\ref{sec3}. Simulation examples are given in Sec.~\ref{sec4}, and conclusions are provided in Sec.~\ref{sec5}.

%%%%%%%%%%%%%%%%%%%%%%%%%%%%%%%%%%%%%%%%%%%%%%%%%%%%%%%%%%%%%%%%%%%%%%%%%%%%%%%%%%%%%
%%%%%%%%%%%%%%%%%%%%%%%%%%%%%%%%%%%%%%%%%%%%%%%%%%%%%%%%%%%%%%%%%%%%%%%%%%%%%%%%%%%%%
\section{Preliminary} \label{sec2}

%%%%%%%%%%%%%%%%%%%%%%%%%%%%%%%%%%%%%%%%%%%%%%%%%%%%%%%%%%%%%%%%%%%%%%%%%%%%%%%%%%%%%
\subsection{Problem Statement} \label{sec2.1}

Consider a passive IROD scenario involving an observer and a target, with orbital states
${\boldsymbol{x}_\Delta}(t)=[{\boldsymbol{r}_\Delta}(t);{\boldsymbol{v}_\Delta}(t)]\in\mathbb{R}^6$, $\Delta\in\{O,T\}$, governed by
\begin{equation}\label{eq3}
    {\boldsymbol{\dot x}_\Delta}(t)=\boldsymbol{f}(t,{\boldsymbol{x}_\Delta}(t)),
\end{equation}
where $\boldsymbol{f}$ represents the nonlinear orbital dynamics. Assuming short-range, maneuver-free motion, the same dynamics are used for both spacecraft. The state at time $t$ is then written as ${\boldsymbol{x}_\Delta}(t)=\boldsymbol{F}({\boldsymbol{x}_\Delta}(t_0),t_0,t)$, where $\boldsymbol{F}$ is the nonlinear flow map associated with Eq.~\eqref{eq3}.

Assuming angle-only sensing, the nominal measurement is the LOS unit vector
\begin{equation}\label{eq5}
    \boldsymbol{\tilde l}(t)=\frac{{\boldsymbol{r}_T}(t)-{\boldsymbol{r}_O}(t)}
    {\|{\boldsymbol{r}_T}(t)-{\boldsymbol{r}_O}(t)\|}\in\mathbb{R}^3 .
\end{equation}
The goal of IROD is to estimate the target state, or equivalently the relative state, from a set of LOS measurements, with the initial epoch aligned with the first observation. Let $\boldsymbol{\varepsilon}(t)\in\mathbb{R}^3$ denote the LOS measurement error. According to the QUEST model \cite{Shuster1981}, $\mathbb{E}\{\boldsymbol{\varepsilon}(t)\}=\mathbf{0}_{3\times1}$ and
\begin{equation}\label{eq7}
    \mathbb{E}\{\boldsymbol{\varepsilon}(t)\boldsymbol{\varepsilon}^T(t)\}
    =\frac{\sigma^2}{2}\left[\boldsymbol{I}_3-\boldsymbol{\tilde l}(t)\boldsymbol{\tilde l}^T(t)\right]
    \equiv \boldsymbol{R}(t),
\end{equation}
where $\sigma$ is the sensor noise standard deviation (STD). The measured LOS vector is then written as
\begin{equation}\label{eq8}
    \boldsymbol{l}(t)=\boldsymbol{\tilde l}(t)+\boldsymbol{\varepsilon}(t)
    =[l_x(t),l_y(t),l_z(t)]^T \in \mathbb{R}^3 .
\end{equation}
For brevity, time-dependent variables are abbreviated hereafter, e.g., $\boldsymbol{l}_i\equiv\boldsymbol{l}(t_i)$.

%%%%%%%%%%%%%%%%%%%%%%%%%%%%%%%%%%%%%%%%%%%%%%%%%%%%%%%%%%%%%%%%%%%%%%%%%%%%%%%%%%%%%
\subsection{High-Order Taylor Polynomial} \label{sec2.2}

Let $\delta {\boldsymbol{x}_0} = {\boldsymbol{x}_T}({t_0}) - {\boldsymbol{x}_O}({t_0})$ denote the initial relative state. The relative state at epoch $t_i$ can be approximated by a high-order Taylor expansion as
\begin{equation} \label{eq9}
    \delta {\boldsymbol{x}_i}
    \equiv \delta \boldsymbol{x}({t_i})
    = {\boldsymbol{x}_T}(t_i) - {\boldsymbol{x}_O}(t_i)
    \approx {{\cal T}_{\delta {\boldsymbol{x}_i}}}(\delta {\boldsymbol{x}_0})
    = \sum\limits_{p = 1}^N {\frac{1}{{p!}}\phi _{({t_0},t_i)}^{i,{k_1} \cdots {k_p}}\delta x_0^{{k_1}} \cdots \delta x_0^{{k_p}}} \,,
\end{equation}
where $N$ is the truncation order and $\phi _{({t_0},t_i)}^{i,{k_1} \cdots {k_p}}$ denotes the coefficient of the $p$th-order term. No constant term appears because $\delta {\boldsymbol{x}_i}=\mathbf{0}$ when $\delta {\boldsymbol{x}_0}=\mathbf{0}$. Compared with linearized dynamics, the high-order polynomial captures richer nonlinear behaviors and improves observability, while remaining much more efficient than numerical propagation.

The polynomial coefficients in Eq.~\eqref{eq9} can be obtained using variational techniques such as STTs \cite{Park2006, Zhou2026TDSTT}, or algorithmic differentiation frameworks such as differential algebra (DA) \cite{Armellin2009} and jet transport (JT) \cite{PerezPalau2015}. Unlike STT-based methods, which require explicit derivation of high-order variational equations, DA and JT generate these coefficients automatically through algebraic differentiation. In this work, DA is adopted because it enables flexible and efficient computation of high-order Taylor expansions without symbolic derivation.

%%%%%%%%%%%%%%%%%%%%%%%%%%%%%%%%%%%%%%%%%%%%%%%%%%%%%%%%%%%%%%%%%%%%%%%%%%%%%%%%%%%%%
\subsection{Orbit Refinement Approaches} \label{sec2.3}

Orbit refinement is used in this work only to assess the quality of the IROD solution, rather than as a methodological contribution. Since a better initial estimate generally leads to faster convergence, lower computational cost, and higher refinement accuracy, the refinement performance provides an indirect but quantitative indicator of IROD quality. Two refinement methods are considered for this purpose: nonlinear least squares (NLS) \cite{Crassidis2011}, which iteratively minimizes angular residuals, and optimal linear orbit determination (OLOD) \cite{Sinclair2020}, which solves a linearized cross-product formulation analytically. Their performance under different IROD outputs will be compared in Sec.~\ref{sec4}.

%%%%%%%%%%%%%%%%%%%%%%%%%%%%%%%%%%%%%%%%%%%%%%%%%%%%%%%%%%%%%%%%%%%%%%%%%%%%%%%%%%%%%
%%%%%%%%%%%%%%%%%%%%%%%%%%%%%%%%%%%%%%%%%%%%%%%%%%%%%%%%%%%%%%%%%%%%%%%%%%%%%%%%%%%%%
\section{Methodology} \label{sec3}

%%%%%%%%%%%%%%%%%%%%%%%%%%%%%%%%%%%%%%%%%%%%%%%%%%%%%%%%%%%%%%%%%%%%%%%%%%%%%%%%%%%%%
\subsection{Formulation of the Initial Relative Orbit Determination Problem} \label{sec3.1}

%%%%%%%%%%%%%%%%%%%%%%%%%%%%%%%%%%%%%%%%%%%%%%%%%%%%%%%%%%%%%%%%%%%%%%%%%%%%%%%%%%%%%
\subsubsection{Measurement-Residual Objective} \label{sec3.1.1}
In angles-only IROD, the measurement residual can be modeled either by directly comparing the observed and predicted LOS directions or by enforcing the collinearity constraint between the relative position and the measured LOS. In this work, the latter is adopted because the resulting cross-product residual is algebraically simple and integrates naturally with the polynomial approximation in Eq.~\eqref{eq9}. The corresponding objective function is written as
\begin{equation} \label{eq10}
    \mathop {\min }\limits_{\delta {\boldsymbol{x}_0}} \,\,\,J = \sum\limits_i {{{[{{({\boldsymbol{l}_i} \times \delta {\boldsymbol{r}_i})}^T}({\boldsymbol{l}_i} \times \delta {\boldsymbol{r}_i})]}^{\frac{n}{2}}}} = \sum\limits_i {{{[{{(\boldsymbol{l}_i^ \times \delta {\boldsymbol{r}_i})}^T}_i(\boldsymbol{l}_i^ \times \delta {\boldsymbol{r}_i})]}^{\frac{n}{2}}}} = \sum\limits_i {{{\left\| {\boldsymbol{l}_i^ \times \delta {\boldsymbol{r}_i}} \right\|}^n}} \,,
\end{equation}
where $\delta \boldsymbol{r}_i \equiv \delta \boldsymbol{r}(t_i)$ is the position component of $\delta \boldsymbol{x}_i$, $n$ is the residual order, and $\boldsymbol{l}_i^\times$ is the matrix form of the cross-product operator. 

Despite its advantages, the cross-product formulation has two important limitations. First, because $\boldsymbol{l}_i^\times \mathbf{0}=\mathbf{0}$ for all $i$, the zero relative state trivially yields zero residual and may therefore appear as a spurious local optimum. Second, the unweighted form in Eq.~\eqref{eq10} does not account for the fact that the cross-product residual is heteroscedastic: even under the same angular noise level, its statistical scale depends on the observation geometry and relative distance. As a result, measurements from different epochs should not, in general, contribute equally to the cost. At the same time, direct weighting is nontrivial because LOS measurements intrinsically contain only two DoF, which can make the corresponding weighting matrix rank-deficient or numerically ill-conditioned. These two issues motivate the reduced-order weighting strategy and zero-solution-avoidance mechanism introduced later.

%%%%%%%%%%%%%%%%%%%%%%%%%%%%%%%%%%%%%%%%%%%%%%%%%%%%%%%%%%%%%%%%%%%%%%%%%%%%%%%%%%%%%
\subsubsection{Zero-Avoidance Constraint} \label{sec3.1.2}
\begin{figure}[!h]
	\centering
	\includegraphics[width=0.65\textwidth]{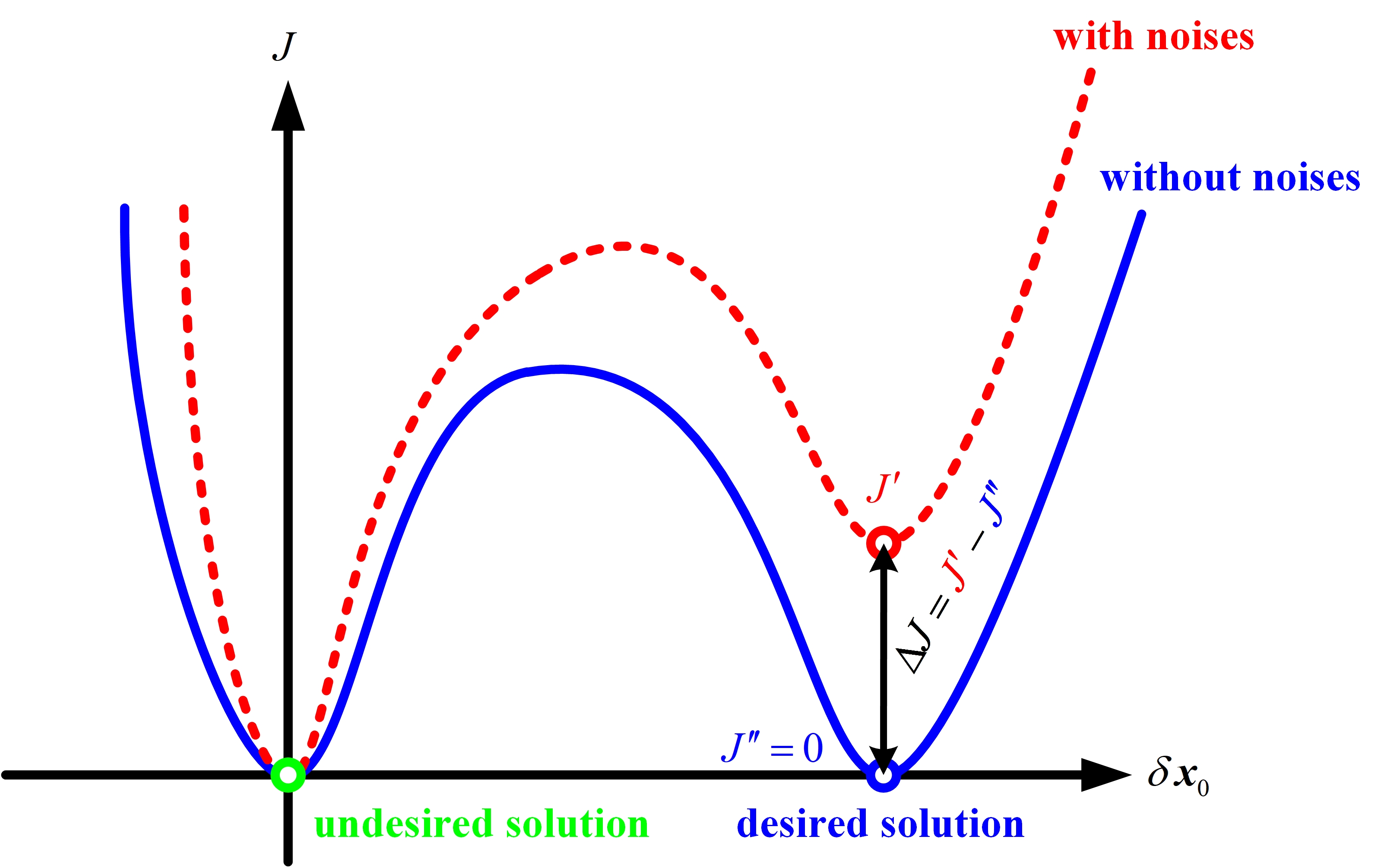}
	\caption{An illustration of the desired and undesired solution.}
	\label{fig1}
\end{figure}

A critical issue of the cross-product-based formulation is the existence of a spurious zero solution, \emph{i.e.}, $\delta \boldsymbol{x}_0=\mathbf{0}_6$. In this case, $\delta \boldsymbol{r}_i=\mathbf{0}_3$ for all $i$, and thus the objective in Eq.~\eqref{eq10} is identically zero because $\boldsymbol{l}_i^\times \mathbf{0}_3=\mathbf{0}_3$. As illustrated in Fig.~\ref{fig1}, this zero solution remains a minimizer regardless of the measurement noise level. In contrast, although the desired solution also gives zero cost in the noise-free case, its objective value increases once measurement noise is present, whereas the zero solution still yields zero residual. As a result, direct minimization of Eq.~\eqref{eq10} becomes strongly biased toward the spurious zero solution under noisy conditions.

To exclude this solution, a zero-avoidance constraint is introduced as $\|\delta \boldsymbol{r}_0\| \ge \delta$, where $\delta>0$ is a prescribed threshold. 
Only the position component is constrained, because identical velocities are physically admissible, whereas coincident positions are not.
The resulting constrained optimization problem is
\begin{equation} \label{eq13}
    \begin{array}{cl}
    \displaystyle \min_{\delta \boldsymbol{x}_0} & J=\sum_i \|\boldsymbol{l}_i^\times \delta \boldsymbol{r}_i\|^n \\
    \text{s.t.} & \|\boldsymbol{H}\delta \boldsymbol{x}_0\| \ge \delta
    \end{array} \,,
\end{equation}
with $\boldsymbol{H}=\left[\boldsymbol{I}_3,\,\mathbf{0}_{3\times3}\right]\in\mathbb{R}^{3\times6}$.

However, Eq.~\eqref{eq13} still has two drawbacks. First, both the objective and the constraint are nonlinear, and evaluating $\delta \boldsymbol{r}_i$ through numerical propagation leads to a computationally expensive nonlinear program. Second, the threshold $\delta$ is user-defined and difficult to choose: a small value may fail to avoid the zero solution, whereas a large value may overly distort the feasible set and degrade the estimation accuracy.

%%%%%%%%%%%%%%%%%%%%%%%%%%%%%%%%%%%%%%%%%%%%%%%%%%%%%%%%%%%%%%%%%%%%%%%%%%%%%%%%%%%%%
\subsubsection{Recursive Polynomial Optimization} \label{sec3.1.3}

The constrained IROD problem in Eq.~\eqref{eq13} can be solved using an existing RPO framework \cite{Pavanello2024IEEE,zhou2025maneuver}. Since the RPO methodology itself is not the focus of this work, only the problem-specific implementation is summarized below.

First, by replacing the numerically propagated relative position with its high-order Taylor approximation, one has $\delta {\boldsymbol{r}_i} \approx {{\cal T}_{\delta {\boldsymbol{r}_i}}}(\delta {\boldsymbol{x}_0})$, where ${{\cal T}_{\delta {\boldsymbol{r}_i}}}$ is the position component of Eq.~\eqref{eq9}. Then, around a current estimate $\delta \hat{\boldsymbol{x}}_0$, the polynomial model is locally linearized as
\begin{equation} \label{eq19}
    \boldsymbol{l}_i^\times \delta \boldsymbol{r}_i
    \approx \boldsymbol{A}_i \Delta \boldsymbol{x}_0 + \boldsymbol{b}_i \,,
\end{equation}
where $\Delta \boldsymbol{x}_0 = \delta \boldsymbol{x}_0-\delta \hat{\boldsymbol{x}}_0$, $\boldsymbol{A}_i=\boldsymbol{l}_i^\times \left.\nabla {{\cal T}_{\delta \boldsymbol{r}_i}}\right|_{\delta \hat{\boldsymbol{x}}_0}$ and $\boldsymbol{b}_i=\boldsymbol{l}_i^\times {{\cal T}_{\delta \boldsymbol{r}_i}}(\delta \hat{\boldsymbol{x}}_0)$.

Introducing auxiliary variables $\chi_i$ and linearizing the zero-avoidance constraint at $\delta \hat{\boldsymbol{x}}_0$ yields the following power cone optimization (PCO) subproblem:
\begin{equation} \label{eq24}
    \begin{array}{cl}
    \displaystyle \min_{\Delta \boldsymbol{x}_0,\,\chi_i} &  J=\sum_i \chi_i \\
    \text{s.t.} & \|\boldsymbol{A}_i\Delta \boldsymbol{x}_0+\boldsymbol{b}_i\|^n \le \chi_i \\
    & (\boldsymbol{H}\delta \hat{\boldsymbol{x}}_0)^T
    \boldsymbol{H}(\delta \hat{\boldsymbol{x}}_0+\Delta \boldsymbol{x}_0)\ge \delta 
    \end{array} \,,
\end{equation}
When $n=2$, Eq.~\eqref{eq24} reduces to a very common second-order cone optimization problem.

At each iteration, Eq.~\eqref{eq24} is solved with MOSEK, and the estimate is updated as $\delta \hat{\boldsymbol{x}}_0 \leftarrow \delta \hat{\boldsymbol{x}}_0 + \Delta \boldsymbol{x}_0^*$, 
where $\Delta \boldsymbol{x}_0^*$ is the optimal increment. The procedure is repeated until $\|\Delta \boldsymbol{x}_0^*\| \le \eta$. The main computational advantage of the RPO is that, by locally linearizing the polynomial and convexifying the zero-avoidance constraint, each iteration reduces to a convex optimization subproblem.

%%%%%%%%%%%%%%%%%%%%%%%%%%%%%%%%%%%%%%%%%%%%%%%%%%%%%%%%%%%%%%%%%%%%%%%%%%%%%%%%%%%%%
\subsection{Reduced-Order Weighting Strategy} \label{sec3.2}

The objective in Eq.~\eqref{eq10} assigns equal weights to all cross-product residuals. This is generally suboptimal, because the residual statistics depend not only on the sensor noise level but also on the observation geometry and the relative range. A natural way to introduce weighting is therefore to derive the covariance of the cross-product residual and use its inverse as the weighting matrix, as in Ref.~\cite{Sinclair2020}.

Define the residual at epoch $t_i$ as $\boldsymbol{e}_i \equiv \boldsymbol{l}_i^\times \delta \boldsymbol{r}_i \in \mathbb{R}^3$.
At the true solution, $\tilde{\boldsymbol{l}}_i^\times \delta \boldsymbol{r}_i=\mathbf{0}$, and with $\boldsymbol{l}_i=\tilde{\boldsymbol{l}}_i+\boldsymbol{\varepsilon}_i$, one has
\begin{equation} \label{eq:w2}
    \boldsymbol{e}_i
    = \boldsymbol{\varepsilon}_i^\times \delta \boldsymbol{r}_i
    = - \delta \boldsymbol{r}_i^\times \boldsymbol{\varepsilon}_i .
\end{equation}
Using the QUEST measurement model in Eq.~\eqref{eq7}, the covariance of the conventional three-dimensional residual is
\begin{equation} \label{eq:w3}
    \boldsymbol{Q}_i
    \equiv \mathbb{E}\{\boldsymbol{e}_i \boldsymbol{e}_i^T\}
    = \delta \boldsymbol{r}_i^\times \boldsymbol{R}_i \delta \boldsymbol{r}_i^{\times T}.
\end{equation}
A conventional weighted formulation would then use $\boldsymbol{Q}_i^{-1}$ as the weighting matrix. However, this covariance is intrinsically singular: LOS measurements contain only two independent DoF, whereas $\boldsymbol{Q}_i \in \mathbb{R}^{3\times3}$ is constructed in a redundant three-dimensional space. Hence, $\boldsymbol{Q}_i$ is rank-deficient. In Ref.~\cite{Sinclair2020}, this issue is handled numerically by replacing the zero eigenvalue with a small positive constant. While effective in practice, such regularization remains a numerical remedy rather than a structural resolution.

To remove this intrinsic singularity, a reduced-order weighting strategy is introduced here. Instead of weighting the redundant three-dimensional residual directly, the residual is projected onto the two-dimensional tangent subspace of the LOS. Let $\boldsymbol{E}_i\in\mathbb{R}^{3\times2}$ be an orthonormal basis satisfying
\begin{equation} \label{eq:w4}
    \boldsymbol{E}_i^T\boldsymbol{E}_i=\boldsymbol{I}_2,
    \quad
    \boldsymbol{E}_i^T\boldsymbol{l}_i=\mathbf{0},
\end{equation}
and define the reduced residual as
\begin{equation} \label{eq:w5}
    \bar{\boldsymbol{e}}_i
    = \boldsymbol{E}_i^T \boldsymbol{e}_i
    = \boldsymbol{E}_i^T \boldsymbol{l}_i^\times \delta \boldsymbol{r}_i
    \in \mathbb{R}^2 .
\end{equation}
Since $\boldsymbol{l}_i^\times \delta \boldsymbol{r}_i$ is always orthogonal to $\boldsymbol{l}_i$, this projection removes only the redundant direction and preserves all informative components of the cross-product residual. The corresponding reduced-order covariance becomes
\begin{equation} \label{eq:w6}
    \bar{\boldsymbol{Q}}_i
    \equiv \mathbb{E}\{\bar{\boldsymbol{e}}_i \bar{\boldsymbol{e}}_i^T\}
    = \boldsymbol{E}_i^T \boldsymbol{Q}_i \boldsymbol{E}_i
    \in \mathbb{R}^{2\times2},
\end{equation}
which is nonsingular as long as the observation geometry is not degenerate. The reduced-order weighting matrix is then naturally defined as $\bar{\boldsymbol{W}}_i = \bar{\boldsymbol{Q}}_i^{-1}$.

Let $\bar{\boldsymbol{U}}_i$ be a factorization of $\bar{\boldsymbol{W}}_i$ such that $\bar{\boldsymbol{U}}_i^T \bar{\boldsymbol{U}}_i = \bar{\boldsymbol{W}}_i$.
Then, the reduced-order weighted objective for a general residual order $n$ is defined as
\begin{equation} \label{eq:w9}
    J_w
    = \sum_i
    \left\|
    \bar{\boldsymbol{U}}_i \boldsymbol{E}_i^T \boldsymbol{l}_i^\times \delta \boldsymbol{r}_i
    \right\|^n .
\end{equation}
When $n=2$, Eq.~\eqref{eq:w9} reduces to the standard Mahalanobis quadratic form,
\begin{equation} \label{eq:w10}
    J_w
    = \sum_i
    \bar{\boldsymbol{e}}_i^T \bar{\boldsymbol{W}}_i \bar{\boldsymbol{e}}_i .
\end{equation}
This reduced-order formulation is the proposed weighting scheme. It is consistent with the two-DoF nature of LOS measurements and avoids constructing an artificial information direction in the redundant three-dimensional space.

A useful closed-form approximation can be further obtained under the QUEST noise model and small-angle assumptions. Let
\begin{equation} \label{eq:w11}
    \delta \boldsymbol{r}_i \approx \rho_i \tilde{\boldsymbol{l}}_i,
    \quad
    \rho_i = \|\delta \boldsymbol{r}_i\| .
\end{equation}
Substituting Eq.~\eqref{eq7} and Eq.~\eqref{eq:w11} into Eq.~\eqref{eq:w6} yields
\begin{equation} \label{eq:w12}
    \bar{\boldsymbol{Q}}_i
    \approx \frac{\sigma^2 \rho_i^2}{2}\boldsymbol{I}_2,
\end{equation}
and thus
\begin{equation} \label{eq:w13}
    \bar{\boldsymbol{W}}_i
    \approx \frac{2}{\sigma^2 \rho_i^2}\boldsymbol{I}_2,
    \quad
    \bar{\boldsymbol{U}}_i
    \approx \sqrt{\frac{2}{\sigma^2 \rho_i^2}}\,\boldsymbol{I}_2.
\end{equation}
Therefore, Eq.~\eqref{eq:w9} reduces to
\begin{equation} \label{eq:w14}
    J_w
    \approx
    \sum_i
    \left(
    \sqrt{\frac{2}{\sigma^2 \rho_i^2}}
    \left\|
    \boldsymbol{E}_i^T \boldsymbol{l}_i^\times \delta \boldsymbol{r}_i
    \right\|
    \right)^n .
\end{equation}
Because $\boldsymbol{l}_i^\times \delta \boldsymbol{r}_i$ already lies in the tangent subspace, the projection does not change its norm, and Eq.~\eqref{eq:w14} can be further simplified as
\begin{equation} \label{eq:w15}
    J_w
    \approx
    \sum_i
    \left(\frac{2}{\sigma^2 \rho_i^2}\right)^{\frac{n}{2}}
    \left\|
    \boldsymbol{l}_i^\times \delta \boldsymbol{r}_i
    \right\|^n .
\end{equation}

In practice, the true range $\rho_i$ is unknown and is replaced by the current estimate, $\hat{\rho}_i= \left\|{\cal T}_{\delta \boldsymbol{r}_i}(\delta \hat{\boldsymbol{x}}_0)\right\|$. 
It should be emphasized that the proposed method is fundamentally a reduced-order covariance weighting scheme constructed in the two-dimensional tangent subspace of the LOS. The inverse-range weighting in Eq.~\eqref{eq:w15} is only a closed-form approximation of this scheme under the QUEST model and small-angle assumptions. It should also be noted that the weighted objective in Eqs.~\eqref{eq:w9} and \eqref{eq:w15} remains compatible with the RPO in Sec.~\ref{sec3.1.3}. Specifically, at each RPO iteration, the weighting matrices are evaluated using the current estimate $\delta \hat{\boldsymbol{x}}_0$ and are therefore treated as constants within the corresponding convex subproblem. As a result, the weighted residual still has the form of a norm of an affine function of the decision variable, so that the weighted optimization problem can be reformulated as a PCO problem and solved within the same RPO framework. 

Table~\ref{tab1} summarizes the core ideas of the three weighting strategies and compares their main advantages and limitations. Among them, the developed reduced-order weighting method is the most theoretically complete, since it eliminates the numerical singularity issue at the structural level while preserving the full statistical meaning of covariance-based weighting. It is therefore expected to provide the highest estimation accuracy among the three methods.

\begin{table}[!h]
\centering
\caption{Comparison of different weighting strategies}
\label{tab1}
\renewcommand{\arraystretch}{1.2}
\setlength{\tabcolsep}{5pt}
\begin{tabular}{p{0.3cm} p{2.1cm} p{4.0cm} p{3.5cm} p{4.5cm}}
\hline\hline
ID & Method & Core idea & Advantage & Limitation \\
\hline
1 & 3-D weighting (Ref.~\cite{Sinclair2020}) 
& Construct the weighting matrix directly in the 3-D cross-product residual space, with numerical treatment for singularity 
& -- 
& The covariance matrix is intrinsically singular; the existing singularity treatment is essentially a numerical remedy and may lead to accuracy loss \\
\hline
2 & Reduced-order weighting (Eq.~\eqref{eq:w9}) 
& Project the residual onto the two-dimensional LOS tangent subspace and construct the weighting matrix in that subspace 
& Eliminates the singularity issue at the structural level and is consistent with the two-DoF nature of LOS measurements 
& Requires an additional matrix factorization step compared with the simplified formulation \\
\hline
3 & Simplified reduced-order weighting (Eq.~\eqref{eq:w15}) 
& Further approximate the reduced-order weighting by a closed-form scalar weight 
& Simple to implement and free of singularity 
& Involves the strongest approximation and therefore suffers the largest accuracy loss \\
\hline\hline
\end{tabular}
\end{table}

%%%%%%%%%%%%%%%%%%%%%%%%%%%%%%%%%%%%%%%%%%%%%%%%%%%%%%%%%%%%%%%%%%%%%%%%%%%%%%%%%%%%%
\subsection{Adaptive Threshold Selection} \label{sec3.3}

Although the RPO in Sec.~\ref{sec3.1.3} greatly improves computational efficiency, two practical issues remain unresolved: the zero-avoidance threshold $\delta$ in Eq.~\eqref{eq13} is still user-defined, and the RPO procedure requires a nonzero initial guess. These two issues are closely coupled in the present problem, because the zero solution is a spurious local optimum of the cross-product objective (as discussed in Sec.~\ref{sec3.1.2}). 
As a result, using the zero vector as the initial guess is not feasible, and an excessively large threshold may exclude the desired optimum from the feasible domain.

To address this issue, an adaptive thresholding strategy is introduced herein. Note that this adaptive thresholding strategy applies to both the unweighted and weighted formulations in Sec.~\ref{sec3.1} and Sec.~\ref{sec3.2}. Let $k$ denote the index of the adaptive threshold iteration, and let $\delta^{(k)}$ be the threshold used at the $k$th trial. For a given $\delta^{(k)}$, the constrained problem in Eq.~\eqref{eq13} is solved first. The resulting solution is then used as the initial guess for the corresponding unconstrained refinement problem. For the unweighted formulation, this problem is written as
\begin{equation} \label{eq26}
    \begin{array}{cl}
    \displaystyle \min_{\Delta \boldsymbol{x}_0,\,\chi_i} &  J=\sum_i \chi_i \\
    \text{s.t.} & \|\boldsymbol{A}_i\Delta \boldsymbol{x}_0+\boldsymbol{b}_i\|^n \le \chi_i \\
    \end{array} \,,
\end{equation}
For the weighted formulation, the same constrained-to-unconstrained transition is applied, with the objective replaced by its weighted counterpart.

The rationale behind this constrained-to-unconstrained transition is to use the unconstrained refinement as a diagnostic test for the effectiveness of the current threshold. The constrained solve first forces the iterate away from the spurious zero solution, but this alone does not guarantee that the obtained point lies in the attraction region of the desired solution. If the threshold is still too small, the constrained solution may remain inside the attraction region of the zero solution, in which case the subsequent unconstrained refinement will collapse back to zero. By contrast, if the unconstrained refinement no longer returns to the zero solution, then the current threshold is regarded as sufficiently large. In this sense, the proposed procedure searches for the smallest effective threshold that moves the iterate out of the zero-solution attraction region without introducing an unnecessarily restrictive feasible set.

After solving the unconstrained problem, the solution is classified as a zero solution if $\|\delta \boldsymbol{r}_0\| \le \varepsilon$, where $\varepsilon$ is a prescribed small positive constant. If this condition is satisfied, then the current threshold is deemed insufficient, indicating that the constrained solution has not been pushed far enough away from the zero-solution attraction region. Otherwise, the current threshold is accepted. In this way, the threshold is not prescribed a priori, but adaptively increased until it becomes large enough to steer the solution away from the undesired zero solution.

The adaptive procedure is initialized from a minimum threshold $\delta^{(0)}=\delta_{\min}$. For each threshold trial, the initial guess of the constrained problem is constructed along the first LOS direction as
\begin{equation} \label{eq27}
    \delta \hat{\boldsymbol{x}}_0^{(k,0)}
    = \left [ \delta^{(k)} \boldsymbol{l}_1 ; \,\mathbf{0}_{3\times1} \right ] \,,
\end{equation}
where the superscript $(k,0)$ denotes the initial guess used for the constrained solve at the $k$th adaptive threshold iteration. This choice is both nonzero and feasible for the constrained problem. If the current threshold fails the above unconstrained test, it is updated as
\begin{equation} \label{eq28}
    \delta^{(k+1)} = \lambda \delta^{(k)}, \quad \lambda > 1,
\end{equation}
and the procedure is repeated. This strategy seeks the smallest effective threshold that avoids the zero solution without introducing an unnecessarily restrictive feasible set.

%%%%%%%%%%%%%%%%%%%%%%%%%%%%%%%%%%%%%%%%%%%%%%%%%%%%%%%%%%%%%%%%%%%%%%%%%%%%%%%%%%%%%
\subsection{Overall Procedure} \label{sec3.4}

Let ${\delta ^{(k)}} \in \mathbb{R}^1$, $\delta \boldsymbol{\hat x}_0^{(k)} \in \mathbb{R}^6$, $\delta \boldsymbol{x}_0^{(k)} \in \mathbb{R}^6$, and $\delta \boldsymbol{\tilde x}_0^{(k)} \in \mathbb{R}^6$ denote the zero-avoidance threshold, the initial guess of the constrained optimization problem, the optimal solution of the constrained optimization problem, and the optimal solution of the unconstrained optimization problem in the \emph{k}th adaptive-threshold loop, respectively. Then, the pseudo-code of the developed ARPO method is given in Algorithm~\ref{alg1}. Starting from $\delta_{\min}$, the method solves the constrained problem and then performs an unconstrained refinement initialized by the constrained solution. If the refined solution still satisfies $\|\boldsymbol H \delta \tilde{\boldsymbol x}_0^{(k)}\| \leq \varepsilon$, the current threshold is regarded as insufficient and is enlarged by a factor $\lambda$. The procedure is repeated until a nonzero refined solution is obtained or the threshold reaches $\delta_{\max}$.

\begin{algorithm*}[!h]
\caption{Pseudo-code of the ARPO method}
\label{alg1}
\begin{algorithmic}[1]
\Require LOS measurements $\boldsymbol l_i$ at epochs $t_i$, observer initial state $\boldsymbol x_O(t_0)$, polynomial order $N$, thresholds $\delta_{\min}$ and $\delta_{\max}$, scaling factor $\lambda$, zero-solution tolerance $\varepsilon$, and RPO tolerance $\eta$
\State Derive the $N$th-order polynomial model using Eq.~\eqref{eq9};  \Comment{using Eq.~\eqref{eq9} within the DA framework}
\State Initialize $k\leftarrow 0$ and $\delta^{(0)}\leftarrow \delta_{\min}$;
\While{$\delta^{(k)} \le \delta_{\max}$}
    \State Initialize the constrained solve with $\delta \hat{\boldsymbol x}_0^{(k)}=[\delta^{(k)}\boldsymbol l_1;\mathbf 0]$; \Comment{Eq.~\eqref{eq27}}
    \State Construct the weighting matrices using the current estimate $\delta \hat{\boldsymbol x}_0^{(k)}$; \Comment{Eqs.~\eqref{eq:w4}-\eqref{eq:w6}}
    \State Solve the constrained problem to obtain $\delta \boldsymbol x_0^{(k)}$; \Comment{Eq.~\eqref{eq24}}
    \State Update the weighting matrices based on $\delta \boldsymbol x_0^{(k)}$; \Comment{Eqs.~\eqref{eq:w4}-\eqref{eq:w6}}
    \State Solve the corresponding unconstrained refinement problem initialized at $\delta \boldsymbol x_0^{(k)}$ to obtain $\delta \tilde{\boldsymbol x}_0^{(k)}$; \Comment{Eq.~\eqref{eq26}}
    \If{$\|\boldsymbol H \delta \tilde{\boldsymbol x}_0^{(k)}\| > \varepsilon$}
        \State Return $\delta \tilde{\boldsymbol x}_0^{(k)}$; \,\,\underline{Exit while loop}
    \Else
        \State Record the constrained solution $\delta \boldsymbol x_0^{(k)}$;
        \State $\delta^{(k+1)} \leftarrow \lambda \delta^{(k)}$ and $k \leftarrow k+1$; \Comment{Eqs.~\eqref{eq27}-\eqref{eq28}}
    \EndIf
\EndWhile
\State Return the recorded candidate with the smallest measurement residual. \Comment{All threshold trials fail}
\end{algorithmic}
\end{algorithm*}

If all threshold trials fail, an empirical fallback strategy is adopted. Specifically, all constrained solutions are recorded, and the final output is selected as the candidate that minimizes the measurement residual,
\begin{equation} \label{eq29}
    k^*=\arg\min_k \sum_i \left\|\boldsymbol l_i-\hat{\boldsymbol l}_i(\delta \boldsymbol x_0^{(k)})\right\|,
\end{equation}
where $\hat{\boldsymbol l}_i(\cdot)$ denotes the predicted LOS corresponding to the candidate solution. Although this strategy does not guarantee the exact desired optimum, it provides a practical nonzero estimate that can still serve as a useful initialization for subsequent orbit refinement.

%%%%%%%%%%%%%%%%%%%%%%%%%%%%%%%%%%%%%%%%%%%%%%%%%%%%%%%%%%%%%%%%%%%%%%%%%%%%%%%%%%%%%
%%%%%%%%%%%%%%%%%%%%%%%%%%%%%%%%%%%%%%%%%%%%%%%%%%%%%%%%%%%%%%%%%%%%%%%%%%%%%%%%%%%%%
\section{Numerical Results} \label{sec4}

%%%%%%%%%%%%%%%%%%%%%%%%%%%%%%%%%%%%%%%%%%%%%%%%%%%%%%%%%%%%%%%%%%%%%%%%%%%%%%%%%%%%%
\subsection{Scenario Design and Analysis} \label{sec4.1}

The simulations are performed under ideal two-body dynamics in a nondimensional inertial frame, such that Eq.~\eqref{eq3} becomes
\begin{equation} \label{eq31}
    \boldsymbol{f}(t,\boldsymbol{x}_\Delta(t)) =
    \left[ 
    \begin{array}{c}
        \boldsymbol{v}_\Delta(t) \\
        -\mu \dfrac{\boldsymbol{r}_\Delta(t)}{\|\boldsymbol{r}_\Delta(t)\|^3}
    \end{array}    
    \right],
    \quad \Delta \in \{O,T\},
\end{equation}
with $\mu=1$. The initial states of the observer and target are listed in Table~\ref{tab2}.

\begin{table}[!h]
	\centering
	\caption{Initial conditions (nondimensional inertial frame)}
	\label{tab2}
	\begin{tabular}{lllllll}
        \hline\hline
        \multirow{2}{*}{Spacecraft} & \multicolumn{3}{c}{Position} & \multicolumn{3}{c}{Velocity} \\ \cline{2-7} 
         & \multicolumn{1}{l}{$x$} & \multicolumn{1}{l}{$y$} & $z$ & \multicolumn{1}{l}{$\dot{x}$} & \multicolumn{1}{l}{$\dot{y}$} & $\dot{z}$ \\ \hline
        Observer & \multicolumn{1}{l}{1} & \multicolumn{1}{l}{0} & 0 & \multicolumn{1}{l}{0} & \multicolumn{1}{l}{1} & 0 \\
        Target & \multicolumn{1}{l}{1.01} & \multicolumn{1}{l}{0.01} & 0 & \multicolumn{1}{l}{0.01} & \multicolumn{1}{l}{1} & 0 \\ \hline\hline
    \end{tabular}
\end{table}

One orbital period ($T=2\pi$) is considered, with 10 equally spaced LOS measurements. Unless otherwise stated, the sensor noise standard deviation is set to $\sigma=10^{-4}$, corresponding to approximately 1 arcmin at the $3\sigma$ level \cite{Kulik2024, Zhou2023AA}. This noise level is used for the nominal case only; its impact will be further examined in the robustness analysis (Sec.~\ref{sec4.3}). A fifth-order Taylor polynomial ($N=5$) is used in the proposed method. 

\begin{figure*}[!h]
	\centering
	\subfigure[]
	{
		\label{fig3a}
		\centering
		\includegraphics[width=0.31\textwidth]{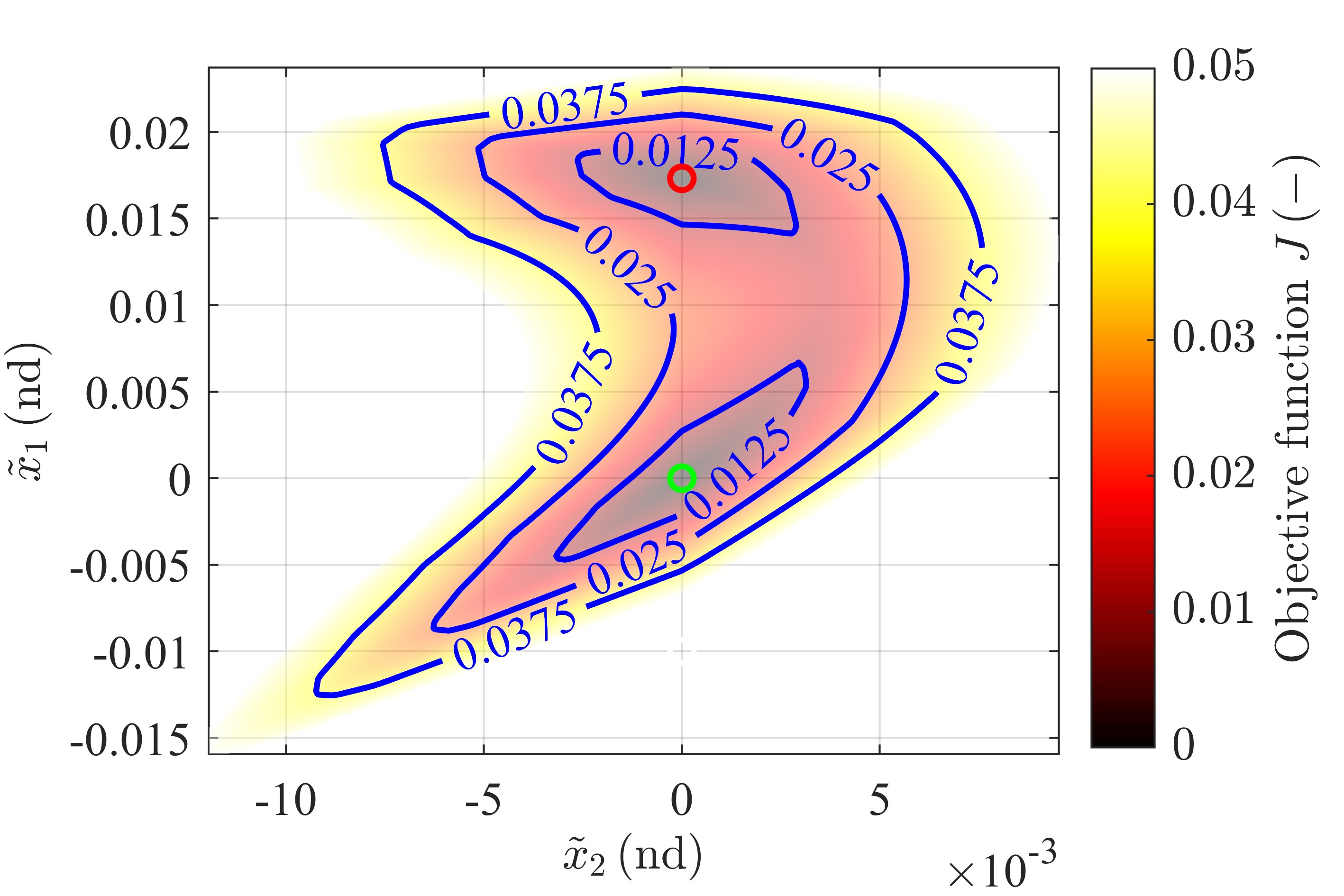}
	}
	\subfigure[]
	{
		\label{fig3b}
		\centering
		\includegraphics[width=0.31\textwidth]{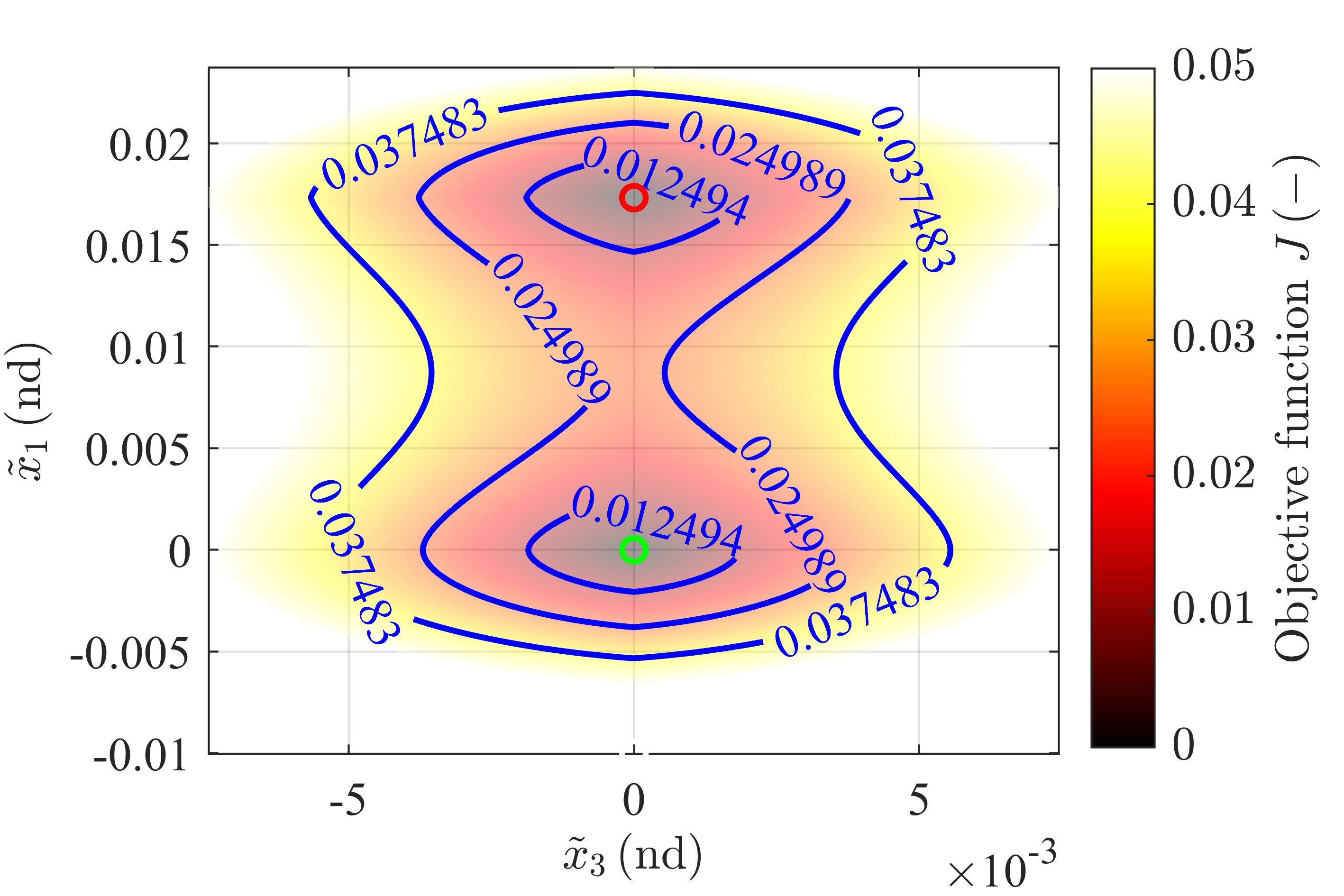}
	}
	\subfigure[]
	{
		\label{fig3c}
		\centering
		\includegraphics[width=0.31\textwidth]{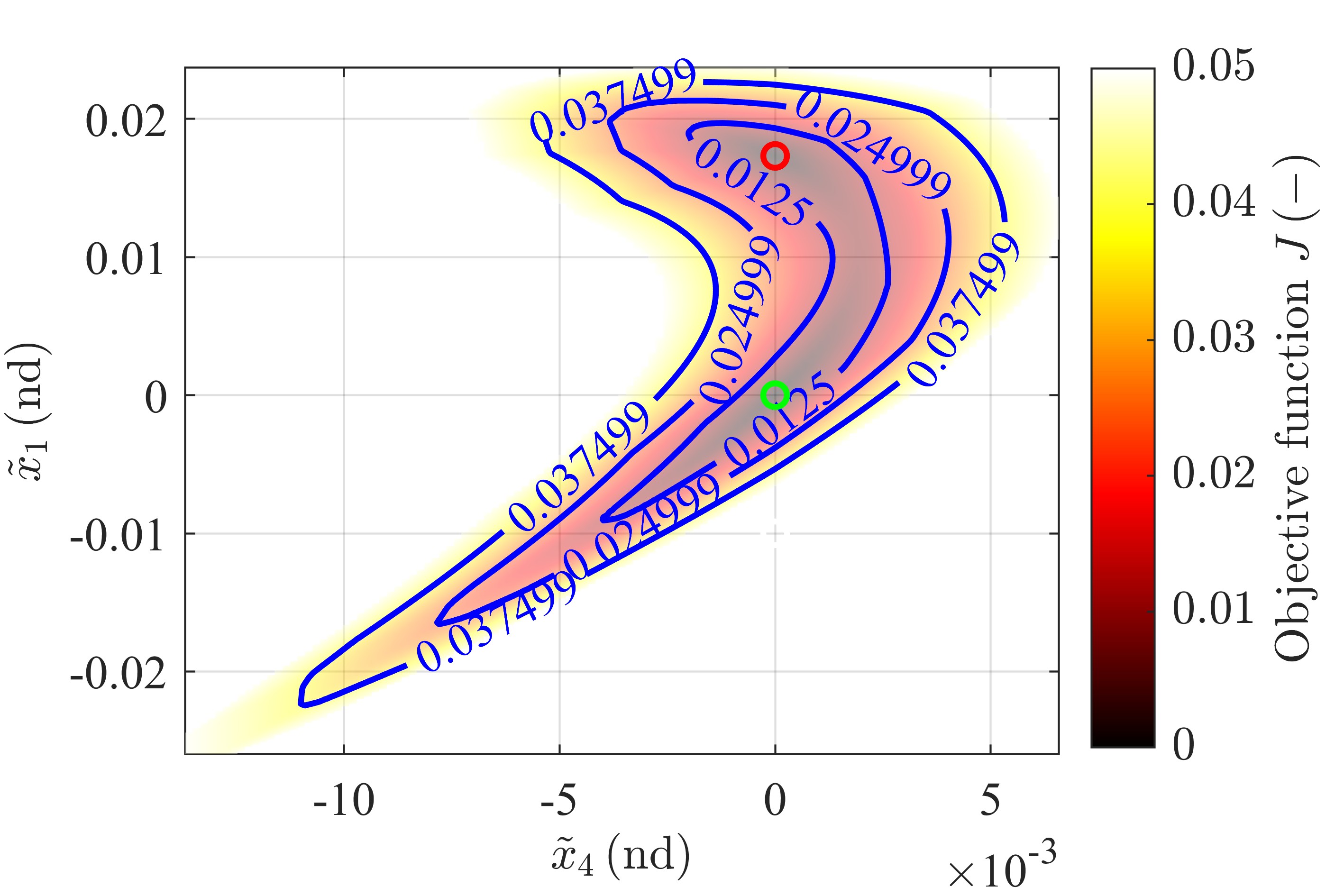}
	}
	\subfigure[]
	{
		\label{fig3d}
		\centering
		\includegraphics[width=0.31\textwidth]{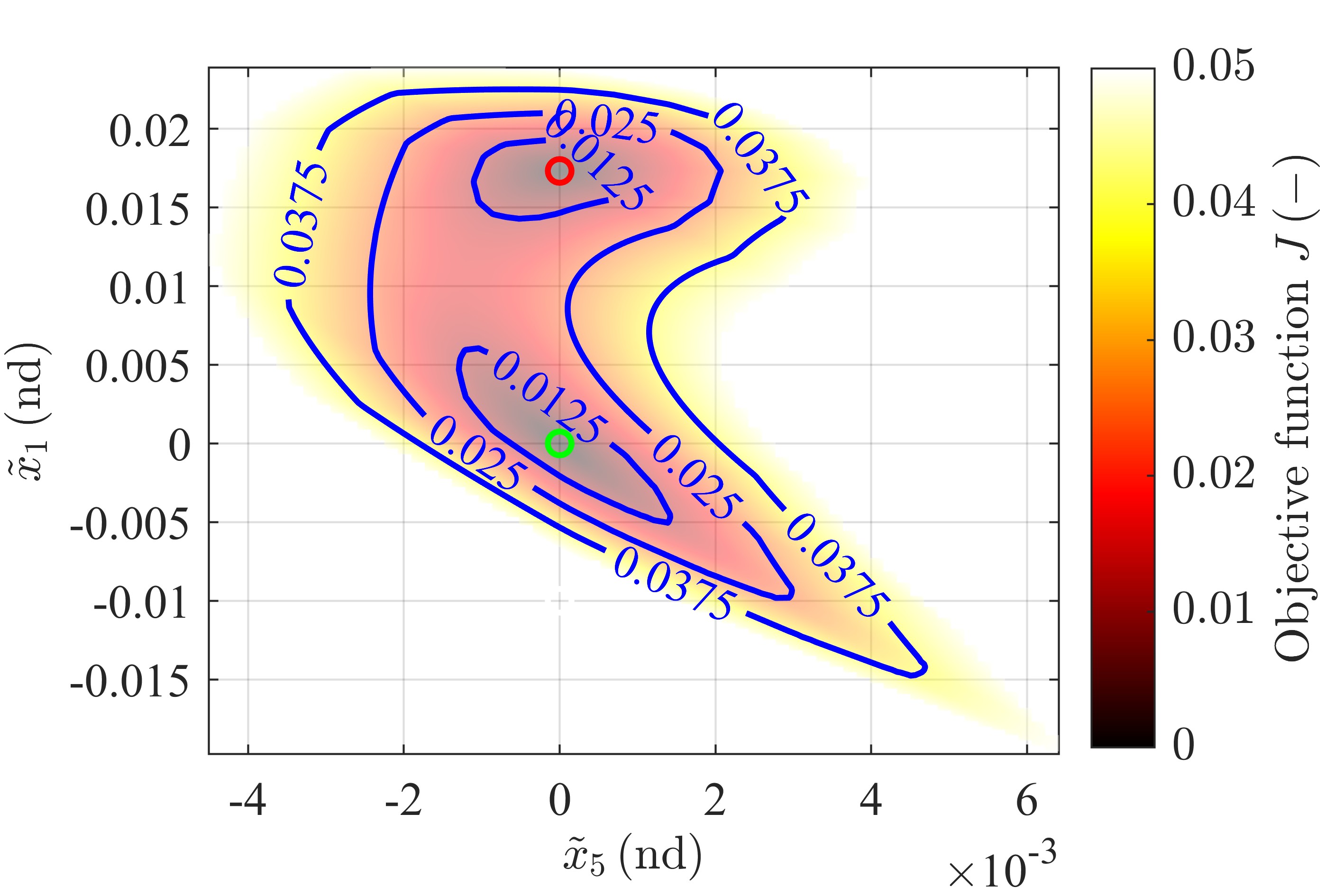}
	}
	\subfigure[]
	{
		\label{fig3e}
		\centering
		\includegraphics[width=0.31\textwidth]{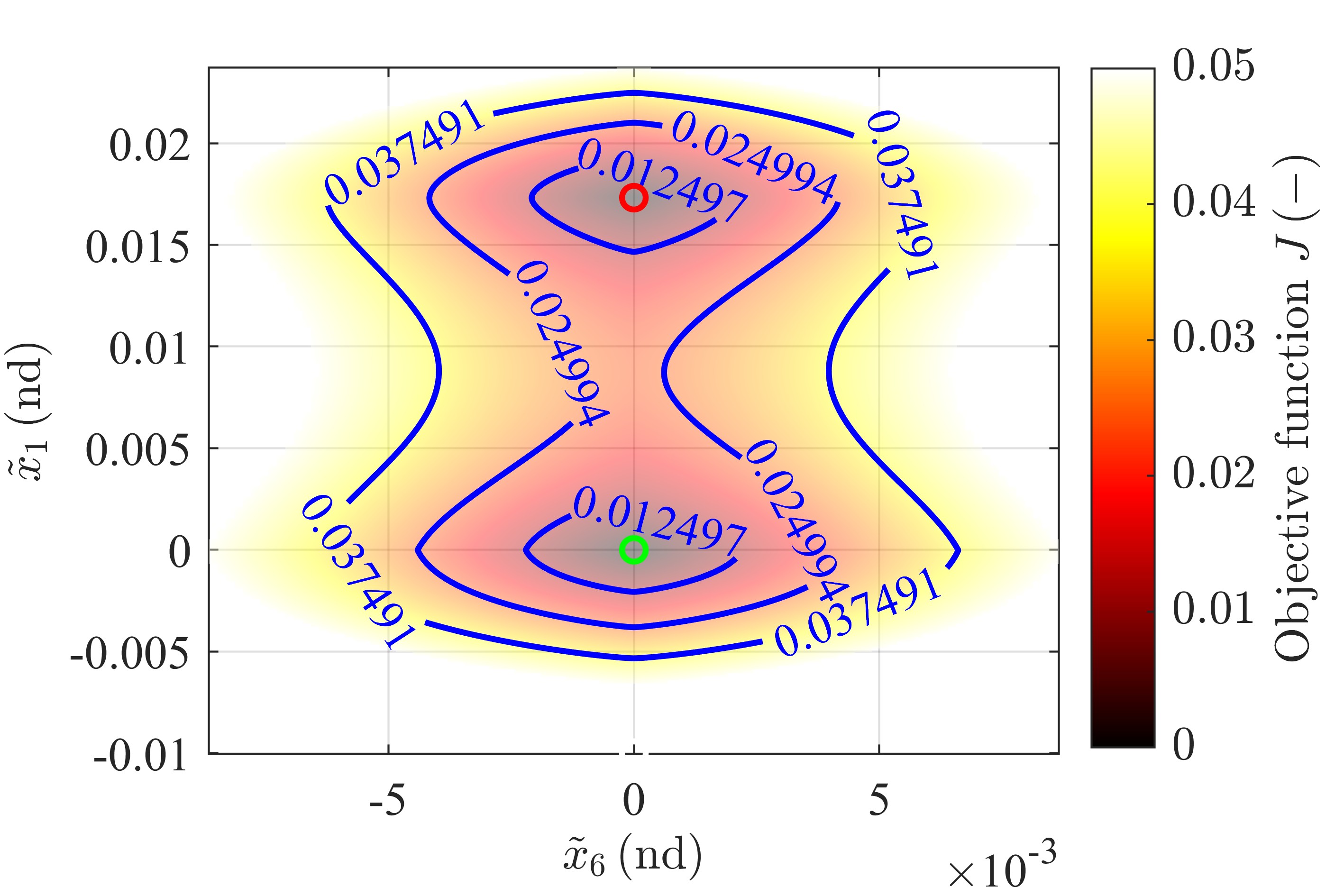}
	}
	\caption{Projected landscape of the unweighted objective function. \subref{fig3a} Projection on $\tilde{x}_1$-$\tilde{x}_2$. \subref{fig3b} Projection on $\tilde{x}_1$-$\tilde{x}_3$. \subref{fig3c} Projection on $\tilde{x}_1$-$\tilde{x}_4$. \subref{fig3d} Projection on $\tilde{x}_1$-$\tilde{x}_5$. \subref{fig3e} Projection on $\tilde{x}_1$-$\tilde{x}_6$.}
	\label{fig3}
\end{figure*}

Figure~\ref{fig3} illustrates the projected landscape of the (unweighted) objective function in Eq.~\eqref{eq10} for the noise-free case with $n$=1. To facilitate visualization, the coordinates are transformed such that the first axis is aligned with the observer-to-target direction at the initial epoch. The figure shows that the objective function exhibits two local minima, corresponding to the desired solution and the spurious zero solution, separated by a ridge. This structure implies that the optimization is sensitive to initialization: if the initial guess lies on the wrong side of the ridge, a gradient-based solver may converge to the undesired zero solution. These results further motivate the introduction of the zero-avoidance constraint in the introduced method.

To simplify the presentation, abbreviations are used throughout this section, as summarized in Table~\ref{tab3}. Although Ref.~\cite{Kulik2024} proposed both pseudoinverse eigenvalue algorithm (PEA) and quadratic eigenvalue algorithm (QEA), only PEA is reported here, since QEA was found to be less accurate in the scenarios considered. Besides PEA, several RPO-based variants are included to isolate the effects of RPO, fixed thresholding, and adaptive thresholding. In particular, P-RPO and P-RPO-$\delta$ assess the roles of RPO and fixed zero-avoidance constraints, respectively, while ARPO is the proposed method. Two ablation variants, L-RPO and U-ARPO, are further introduced to separate the effects of large-threshold initialization and adaptive constrained solving. Finally, OLOD and NLS are used as orbit refinement methods for downstream comparison.

\begin{table*}[!h]
\centering
\caption{Abbreviations for different methods}
\label{tab3}
\begin{tabularx}{\textwidth}{p{1.0cm}p{1.9cm}p{1.3cm}Xp{1.8cm}}
\hline\hline
\multicolumn{2}{l}{ROD} & Abbr. & Description & Reference \\ \hline
\multirow{7}{*}{\makecell[l]{IROD \\stage}} & \multirow{5}{*}{\makecell[l]{Non-adaptive\\ methods}} & PEA & Pseudoinverse eigenvalue algorithm & Ref.~\cite{Kulik2024} \\ \cline{3-5}
& & P-RPO & Single RPO: (a) without the zero-avoidance constraint; (b) using PEA to provide an initial guess to activate the RPO & Eq.~\eqref{eq26} \\ \cline{3-5}
& & P-RPO-$\delta$ & Single RPO: (a) with the zero-avoidance constraint; (b) using PEA to provide an initial guess to activate the RPO & Eq.~\eqref{eq24} \\ \cline{3-5}
& & L-RPO & RPO using a very large zero-avoidance threshold & Eq.~\eqref{eq24} \\ \cline{2-5}
& \multirow{2}{*}{\makecell[l]{Adaptive\\ methods}} & ARPO & Adaptive recursive polynomial optimization & Algorithm~\ref{alg1} \\ \cline{3-5}
& & U-ARPO & ARPO with only unconstrained optimization problem & Eq.~\eqref{eq26} \\ \hline
\multicolumn{2}{l}{\multirow{2}{*}{Orbit refinement stage}} & OLOD & Optimal linear orbit determination & Ref.~\cite{Sinclair2020} \\ \cline{3-5}
\multicolumn{2}{l}{}
& NLS & Nonlinear least squares & Ref.~\cite{Crassidis2011} \\ \hline\hline
\end{tabularx}
\end{table*}

%%%%%%%%%%%%%%%%%%%%%%%%%%%%%%%%%%%%%%%%%%%%%%%%%%%%%%%%%%%%%%%%%%%%%%%%%%%%%%%%%%%%%
\subsection{Performance Analysis on Nominal Case} \label{sec4.2}
Three hundred Monte Carlo (MC) runs are performed for each IROD method. In each run, LOS measurement noise is generated according to the QUEST model, and the same set of pre-generated noisy measurements is used for all methods to ensure a fair comparison. Unless otherwise stated, the user-defined parameters in Algorithm~\ref{alg1} are set as follows: $N=5$, $\eta=10^{-6}$, $\varepsilon=10^{-4}$, $\delta_{\min}=10^{-3}$, $\delta_{\max}=10^{-1}$, and $\lambda=2$. In addition, for P-RPO-$\delta$, the fixed zero-avoidance threshold is set to $\delta=10^{-3}$.

The following nondimensional relative error is used to evaluate both the IROD and refinement results:
\begin{equation} \label{eq33}
    e(\delta {\boldsymbol{x}_0}) = \frac{{\left\| {\delta {\boldsymbol{x}_0} - \delta {{\boldsymbol{\hat x}}_0}} \right\|}}{{\left\| {\delta {\boldsymbol{x}_0}} \right\|}} \,,
\end{equation}
where $\delta {\boldsymbol{x}_0}$ and $\delta {\boldsymbol{\hat x}}_0$ denote the true and estimated relative states at the initial epoch, respectively. The mean relative error (MRE) is defined as the average of Eq.~\eqref{eq33} over the 300 MC runs.

Table~\ref{tab4} compares the accuracy of the non-adaptive methods (unweighted), except for L-RPO, whose results will be discussed separately in Sec.~\ref{sec4.3}. The column ``CR'' denotes the convergence rate of the orbit refinement method. Two failure modes are considered in this work: convergence to the undesired zero solution, which mainly occurs in OLOD, and numerical divergence, which is more common in NLS when inaccurate IROD solutions lead to unreliable gradient information. A run is classified as divergent if the estimated relative position exceeds ten times the true relative distance during any refinement iteration. For the nominal case in Table~\ref{tab4}, the measurement noise is moderate and all IROD methods provide sufficiently accurate initial estimates. Consequently, both OLOD and NLS achieve a convergence rate of 100\% in all runs. The dependence of refinement convergence on IROD accuracy becomes more evident in the robustness analysis presented later.
Table~\ref{tab4} shows that the RPO-based methods significantly improve the IROD accuracy over the baseline PEA and also reduce the number of iterations required by the subsequent refinement methods. For example, compared with PEA, P-RPO reduces the MRE from 1.4566 to \(1.7868\times10^{-3}\) and decreases the average number of OLOD iterations from 6.00 to 2.14. Under this nominal case, the fixed zero-avoidance constraint has little effect on the final results, because the observation noise is low and the PEA initialization is already sufficiently far from the zero solution. Its benefit becomes more apparent in the robustness analysis.

\begin{table*}[!h]
	\centering
	\caption{Accuracy performance of non-adaptive methods (unweighted)}
	\label{tab4}
	\begin{tabular}{lllllll}
    \hline\hline
    \multicolumn{2}{l}{Method} & \multirow{2}{*}{\makecell[l]{Residual \\order $n$}} & \multicolumn{2}{l}{MRE (-)} & \multirow{2}{*}{\makecell[l]{Averaged iteration \\for refinement}} & \multirow{2}{*}{\makecell[l]{CR for \\refinement}} \\ \cline{1-2} \cline{4-5}
    \multicolumn{1}{l}{IROD} & Refinement & & \multicolumn{1}{l}{IROD} & Refinement & & \\ \hline
    \multicolumn{1}{l}{PEA} & OLOD & - & \multicolumn{1}{l}{1.4566} & 1.9644$\times 10^{-3}$ & 6.0000 & 1 \\
    \multicolumn{1}{l}{P-RPO} & OLOD & 1 & \multicolumn{1}{l}{1.7868$\times 10^{-3}$} & 1.9644$\times 10^{-3}$ & 2.1400 & 1 \\
    \multicolumn{1}{l}{P-RPO} & OLOD & 2 & \multicolumn{1}{l}{1.9631$\times 10^{-3}$} & 1.9644$\times 10^{-3}$ & 2.0000 & 1 \\ 
    \multicolumn{1}{l}{P-RPO-$\delta$} & OLOD & 1 & \multicolumn{1}{l}{1.7869$\times 10^{-3}$} & 1.9644$\times 10^{-3}$ & 2.1400 & 1 \\ 
    \multicolumn{1}{l}{P-RPO-$\delta$} & OLOD & 2 & \multicolumn{1}{l}{1.9625$\times 10^{-3}$} & 1.9644$\times 10^{-3}$ & 2.0000 & 1 \\
    \multicolumn{1}{l}{PEA} & NLS & - & \multicolumn{1}{l}{1.4566} & 9.4247$\times 10^{-4}$ & 5.0000 & 1 \\
    \multicolumn{1}{l}{P-RPO} & NLS & 1 & \multicolumn{1}{l}{1.7868$\times 10^{-3}$} & 9.4247$\times 10^{-4}$ & 2.5700 & 1 \\
    \multicolumn{1}{l}{P-RPO} & NLS & 2 & \multicolumn{1}{l}{1.9631$\times 10^{-3}$} & 9.4247$\times 10^{-4}$ & 2.7567 & 1 \\
    \multicolumn{1}{l}{P-RPO-$\delta$} & NLS & 1 & \multicolumn{1}{l}{1.7869$\times 10^{-3}$} & 9.4247$\times 10^{-4}$ & 2.5700 & 1 \\
    \multicolumn{1}{l}{P-RPO-$\delta$} & NLS & 2 & \multicolumn{1}{l}{1.9625$\times 10^{-3}$} & 9.4247$\times 10^{-4}$ & 2.7600 & 1 \\ \hline\hline
    \end{tabular}
\end{table*}

Table~\ref{tab5} reports the computational costs of the non-adaptive methods, except for L-RPO. The four columns correspond to the CPU times of the entire IROD stage, the RPO iterations within IROD, the refinement stage, and the total runtime per MC run. The results show that, within the IROD stage, the RPO iterations account for only a small fraction of the total cost (about 0.02 s), whereas the dominant cost comes from DA-based polynomial propagation (about 0.5 s). Although RPO slightly increases the IROD time, this extra cost is modest, and the improved IROD accuracy significantly reduces the refinement burden. As a result, the total runtime decreases, showing that the RPO-based methods improve both estimation accuracy and overall computational efficiency.

\begin{table*}[!h]
	\centering
	\caption{Computational costs of non-adaptive methods (unweighted)}
	\label{tab5}
	\begin{tabular}{lllllll}
        \hline\hline
        \multicolumn{2}{l}{Method} & \multirow{2}{*}{\makecell[l]{Residual \\order $n$}} & \multicolumn{4}{l}{Averaged CPU time (s)} \\ \cline{1-2} \cline{4-7} 
        \multicolumn{1}{l}{IROD} & Refinement & & \multicolumn{1}{l}{IROD} & \multicolumn{1}{l}{RPO in IROD} & \multicolumn{1}{l}{Refinement} & Total \\ \hline
        \multicolumn{1}{l}{PEA} & OLOD & - & \multicolumn{1}{l}{0.4315} & \multicolumn{1}{l}{-} & \multicolumn{1}{l}{2.2923} & 2.7238 \\
        \multicolumn{1}{l}{P-RPO} & OLOD & 1 & \multicolumn{1}{l}{0.5166} & \multicolumn{1}{l}{0.0177} & \multicolumn{1}{l}{0.8052} & 1.3218 \\
        \multicolumn{1}{l}{P-RPO} & OLOD & 2 & \multicolumn{1}{l}{0.5216} & \multicolumn{1}{l}{0.0172} & \multicolumn{1}{l}{0.7617} & 1.2833 \\
        \multicolumn{1}{l}{P-RPO-$\delta$} & OLOD & 1 & \multicolumn{1}{l}{0.5182} & \multicolumn{1}{l}{0.0197} & \multicolumn{1}{l}{0.8056} & 1.3237 \\
        \multicolumn{1}{l}{P-RPO-$\delta$} & OLOD & 2 & \multicolumn{1}{l}{0.5199} & \multicolumn{1}{l}{0.0204} & \multicolumn{1}{l}{0.7538} & 1.2737 \\
        \multicolumn{1}{l}{PEA} & NLS & - & \multicolumn{1}{l}{0.4304} & \multicolumn{1}{l}{-} & \multicolumn{1}{l}{1.8845} & 2.3149 \\
        \multicolumn{1}{l}{P-RPO} & NLS & 1 & \multicolumn{1}{l}{0.5160} & \multicolumn{1}{l}{0.0177} & \multicolumn{1}{l}{0.9625} & 1.4784 \\
        \multicolumn{1}{l}{P-RPO} & NLS & 2 & \multicolumn{1}{l}{0.5146} & \multicolumn{1}{l}{0.0170} & \multicolumn{1}{l}{1.0293} & 1.5439 \\
        \multicolumn{1}{l}{P-RPO-$\delta$} & NLS & 1 & \multicolumn{1}{l}{0.5191} & \multicolumn{1}{l}{0.0198} & \multicolumn{1}{l}{0.9642} & 1.4833 \\
        \multicolumn{1}{l}{P-RPO-$\delta$} & NLS & 2 & \multicolumn{1}{l}{0.5198} & \multicolumn{1}{l}{0.0204} & \multicolumn{1}{l}{1.0355} & 1.5554 \\ \hline\hline
    \end{tabular}
\end{table*}

Table~\ref{tab6} and Table~\ref{tab7} report the estimation errors and computational costs of ARPO, respectively. 
Although ARPO requires multiple RPO calls, the resulting cost increase is modest. In this nominal case, ARPO attains the same accuracy as the non-adaptive RPO methods, whereas its main advantage becomes evident in the robustness analysis presented in Sec.~\ref{sec4.3}.

\begin{table*}[!h]
	\centering
	\caption{Accuracy performance of the introduced ARPO method (unweighted)}
	\label{tab6}
	\begin{tabular}{llllllll}
    \hline\hline
    \multicolumn{2}{l}{Method} & \multirow{2}{*}{\makecell[l]{Residual \\order $n$}} & \multicolumn{2}{l}{MRE (-)} & \multirow{2}{*}{\makecell[l]{Averaged iteration \\for refinement}} & \multicolumn{2}{l}{CR} \\ \cline{1-2} \cline{4-5} \cline{7-8} 
    \multicolumn{1}{l}{IROD} & Refinement & & \multicolumn{1}{l}{IROD} & Refinement & & \multicolumn{1}{l}{IROD} & Refinement \\ \hline
    \multicolumn{1}{l}{ARPO} & OLOD & 1 & \multicolumn{1}{l}{1.7868$\times 10^{-3}$} & 1.9644$\times 10^{-3}$ & 2.1400 & \multicolumn{1}{l}{1} & 1 \\
    \multicolumn{1}{l}{ARPO} & OLOD & 2 & \multicolumn{1}{l}{1.9631$\times 10^{-3}$} & 1.9644$\times 10^{-3}$ & 2.0000 & \multicolumn{1}{l}{1} & 1 \\
    \multicolumn{1}{l}{ARPO} & NLS & 1 & \multicolumn{1}{l}{1.7868$\times 10^{-3}$} & 9.4247$\times 10^{-4}$ & 2.5700 & \multicolumn{1}{l}{1} & 1 \\
    \multicolumn{1}{l}{ARPO} & NLS & 2 & \multicolumn{1}{l}{1.9631$\times 10^{-3}$} & 9.4247$\times 10^{-4}$ & 2.7567 & \multicolumn{1}{l}{1} & 1 \\ \hline\hline
    \end{tabular}
\end{table*}

\begin{table*}[!h]
	\centering
	\caption{Computational costs of the introduced ARPO method (unweighted)}
	\label{tab7}
	\begin{tabular}{lllllll}
    \hline\hline
    \multicolumn{2}{l}{Method} & \multirow{2}{*}{Residual order $n$} & \multicolumn{4}{l}{Averaged CPU time (s)} \\ \cline{1-2} \cline{4-7} 
    \multicolumn{1}{l}{IROD} & Refinement & & \multicolumn{1}{l}{IROD} & \multicolumn{1}{l}{RPO in IROD} & \multicolumn{1}{l}{Refinement} & Total  \\ \hline
    \multicolumn{1}{l}{ARPO} & OLOD & 1 & \multicolumn{1}{l}{0.5888} & \multicolumn{1}{l}{0.1062} & \multicolumn{1}{l}{0.7979} & 1.3867 \\
    \multicolumn{1}{l}{ARPO} & OLOD & 2 & \multicolumn{1}{l}{0.5870} & \multicolumn{1}{l}{0.1052} & \multicolumn{1}{l}{0.7437} & 1.3307 \\
    \multicolumn{1}{l}{ARPO} & NLS & 1 & \multicolumn{1}{l}{0.5895} & \multicolumn{1}{l}{0.1061} & \multicolumn{1}{l}{0.9563} & 1.5458 \\
    \multicolumn{1}{l}{ARPO} & NLS & 2 & \multicolumn{1}{l}{0.5905} & \multicolumn{1}{l}{0.1060} & \multicolumn{1}{l}{1.0247} & 1.6152 \\ \hline\hline
    \end{tabular}
\end{table*}

In addition to the unweighted scheme, 300 MC runs were also performed for the weighted formulations. In the nominal case considered here, the weighted versions of P-RPO, P-RPO-$\delta$, and ARPO yield identical estimation results; therefore, for brevity, only the ARPO results are reported in Table~\ref{tab8}. 
Table~\ref{tab8} includes the unweighted scheme together with the three weighting strategies listed in Table~\ref{tab1}, namely the 3-D weighting, the reduced-order weighting, and its simplified version. 
For the 3-D weighting strategy, to alleviate the numerical singularity of $\boldsymbol{Q}_i$ (\emph{i.e.}, Eq.~\eqref{eq:w3}), all eigenvalues below $10^{-14}$ are replaced by $10^{-14}$; if $\boldsymbol{Q}_i$ is still numerically singular, the threshold is further increased to $10^{-12}$.
Following Ref.~\cite{Sinclair2020}, both covariance and bias metrics are evaluated. 
Specifically, from the 300 MC solutions, the sample covariance $\boldsymbol{P}\in\mathbb{R}^{6\times6}$ and bias $\boldsymbol{d}\in\mathbb{R}^{6}$ are first computed and partitioned into position and velocity components as
\begin{equation} \label{eq34}
    \begin{array}{*{20}{c}}
    {\boldsymbol{P} = \left[ {\begin{array}{*{20}{c}}
    {{\boldsymbol{P}_{rr}}}&{{\boldsymbol{P}_{rv}}}\\
    {{\boldsymbol{P}_{vr}}}&{{\boldsymbol{P}_{vv}}}
    \end{array}} \right],}&{\boldsymbol{d} = \left[ {\begin{array}{*{20}{c}}
    {{\boldsymbol{d}_r}}\\
    {{\boldsymbol{d}_v}}
    \end{array}} \right]}
    \end{array}
\end{equation}
The covariance metrics are then defined as the square roots of the largest eigenvalues of $\boldsymbol{P}_{rr}$ and $\boldsymbol{P}_{vv}$, denoted by $\varsigma_r$ and $\varsigma_v$, respectively, while the bias metrics are given by the norms $\|\boldsymbol{d}_r\|$ and $\|\boldsymbol{d}_v\|$.

\begin{table*}[!h]
	\centering
	\caption{A comparison between weighted and unweighted schemes (ARPO, $n$=1)}
	\label{tab8}
	\begin{tabular}{lccccc}
    \hline\hline
    \multirow{3}{*}{Weighting strategy} & \multicolumn{5}{c}{IROD results} \\ \cline{2-6}
    & \multicolumn{2}{c}{Bias metrics (nd)} & \multicolumn{2}{c}{Covariance metrics (nd)} & \multirow{2}{*}{\makecell[c]{Averaged CPU\\ time (s)}} \\ \cline{2-5}
    & Position & Velocity & Position & Velocity & \\ \hline
    Unweighted (Eq.~\eqref{eq24}) & 1.3375$\times 10^{-6}$ & 1.2755$\times 10^{-6}$ & 3.1513$\times 10^{-5}$ & 2.0772$\times 10^{-5}$ & 0.1062 \\
    Weighted-1 (Ref.~\cite{Sinclair2020}) & 1.7628$\times 10^{-6}$ & 1.5867$\times 10^{-6}$ & 1.7050$\times 10^{-5}$ & 1.1855$\times 10^{-5}$ & 0.1972 \\
    Weighted-2 (Eq.~\eqref{eq:w9}) & 1.7628$\times 10^{-6}$ & 1.5867$\times 10^{-6}$ & 1.7050$\times 10^{-5}$ & 1.1855$\times 10^{-5}$ & 0.2312 \\ 
    Weighted-3 (Eq.~\eqref{eq:w15}) & 1.4407$\times 10^{-6}$ & 1.3528$\times 10^{-6}$ & 2.0409$\times 10^{-5}$ & 1.3836$\times 10^{-5}$ & 0.1315 \\ \hline\hline                              
    \end{tabular}
\end{table*}

As shown in Table~\ref{tab8}, all three weighted schemes outperform the unweighted one in terms of covariance. Compared with the unweighted scheme, the 3-D and reduced-order weighting strategies reduce the position and velocity covariance metrics from $3.1513\times10^{-5}$ and $2.0772\times10^{-5}$ to $1.7050\times10^{-5}$ and $1.1855\times10^{-5}$, corresponding to reductions of about $45.9\%$ and $42.9\%$, respectively. The simplified weighting strategy also improves the covariance metrics, but to a lesser extent, with reductions of about $35.2\%$ in position and $33.4\%$ in velocity. Among the three weighted schemes, the reduced-order and 3-D weighting methods are nearly identical and both outperform the simplified one. This is because, in the nominal case, the 3-D weighting does not suffer from numerical singularity, whereas the simplified weighting introduces additional approximation error due to model reduction. 
Weighted solutions also show larger bias metrics, consistent with Ref.~\cite{Sinclair2020}.
In terms of computational cost, all weighted schemes are more expensive than the unweighted case, while the simplified weighting is the least expensive among the three weighted methods.

% Figure~\ref{fig4} compares the optimized objective value $J^*$ obtained by ARPO with $n$=1 and the objective value $J'$ evaluated at the desired true solution. In the absence of measurement noise, $J'$ would be zero; in practice, however, it becomes nonzero due to noisy LOS measurements. In both subplots, the blue line represents $J^*=J'$. All samples lie close to this line and in the region $J'>J^*$, indicating that the optimized solution is very close to the true one and that the remaining gap is mainly induced by measurement noise. The observation $J'>J^*$ suggests that, for a given noise realization, the noisy measurements may favor a nearby solution whose objective value is slightly smaller than that of the true state.

% \begin{figure}[!h]
% 	\centering
% 	\subfigure[]
% 	{
% 		\label{fig4a}
% 		\centering
% 		\includegraphics[width=0.45\textwidth]{figures/Typical/objective_J.jpg}
% 	}
% 	\subfigure[]
% 	{
% 		\label{fig4b}
% 		\centering
% 		\includegraphics[width=0.45\textwidth]{figures/Typical/objective_J_w.jpg}
% 	}
% 	\caption{A comparison between the desired and the optimized objective values. \subref{fig4a} Unweighted. \subref{fig4b} Weighted.}
% 	\label{fig4}
% \end{figure}

%%%%%%%%%%%%%%%%%%%%%%%%%%%%%%%%%%%%%%%%%%%%%%%%%%%%%%%%%%%%%%%%%%%%%%%%%%%%%%%%%%%%%
\subsection{Robustness Analysis} \label{sec4.3}

\begin{figure*}[!h]
    \centering
    \subfigure[]{
        \label{fig5a}
        \includegraphics[width=0.45\textwidth]{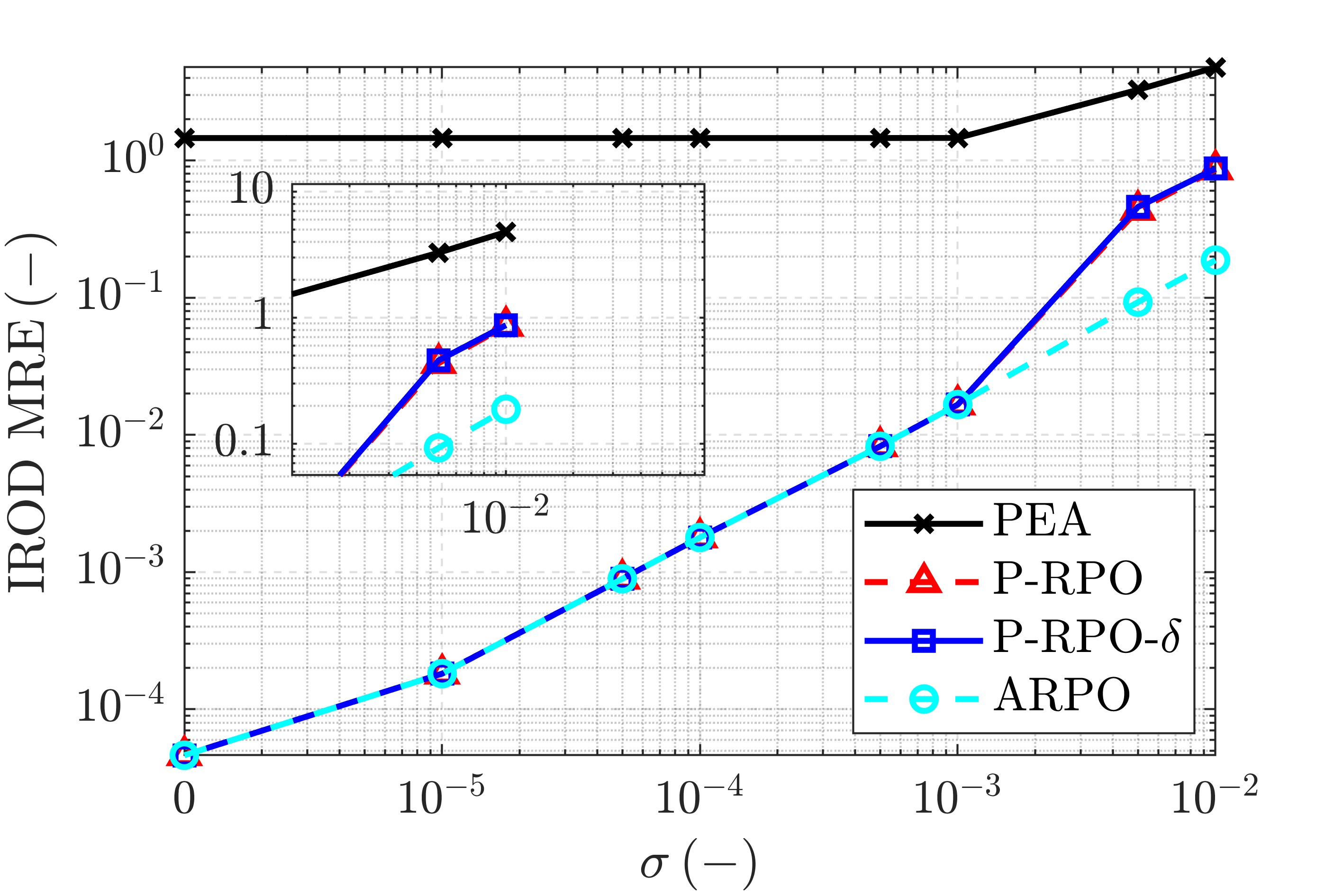}
    }
    \subfigure[]{
        \label{fig5b}
        \includegraphics[width=0.45\textwidth]{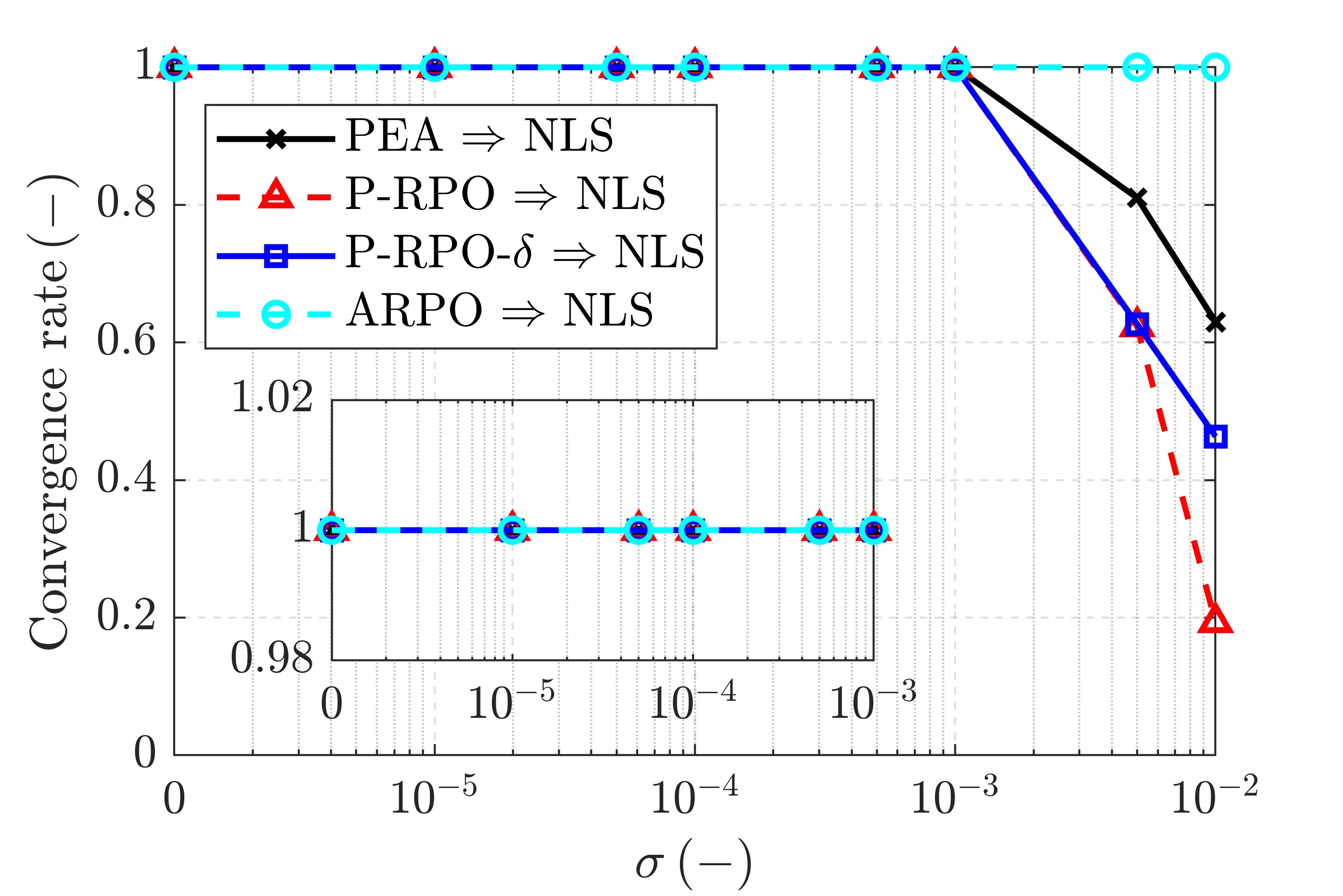}
    }
    \subfigure[]{
        \label{fig5c}
        \includegraphics[width=0.45\textwidth]{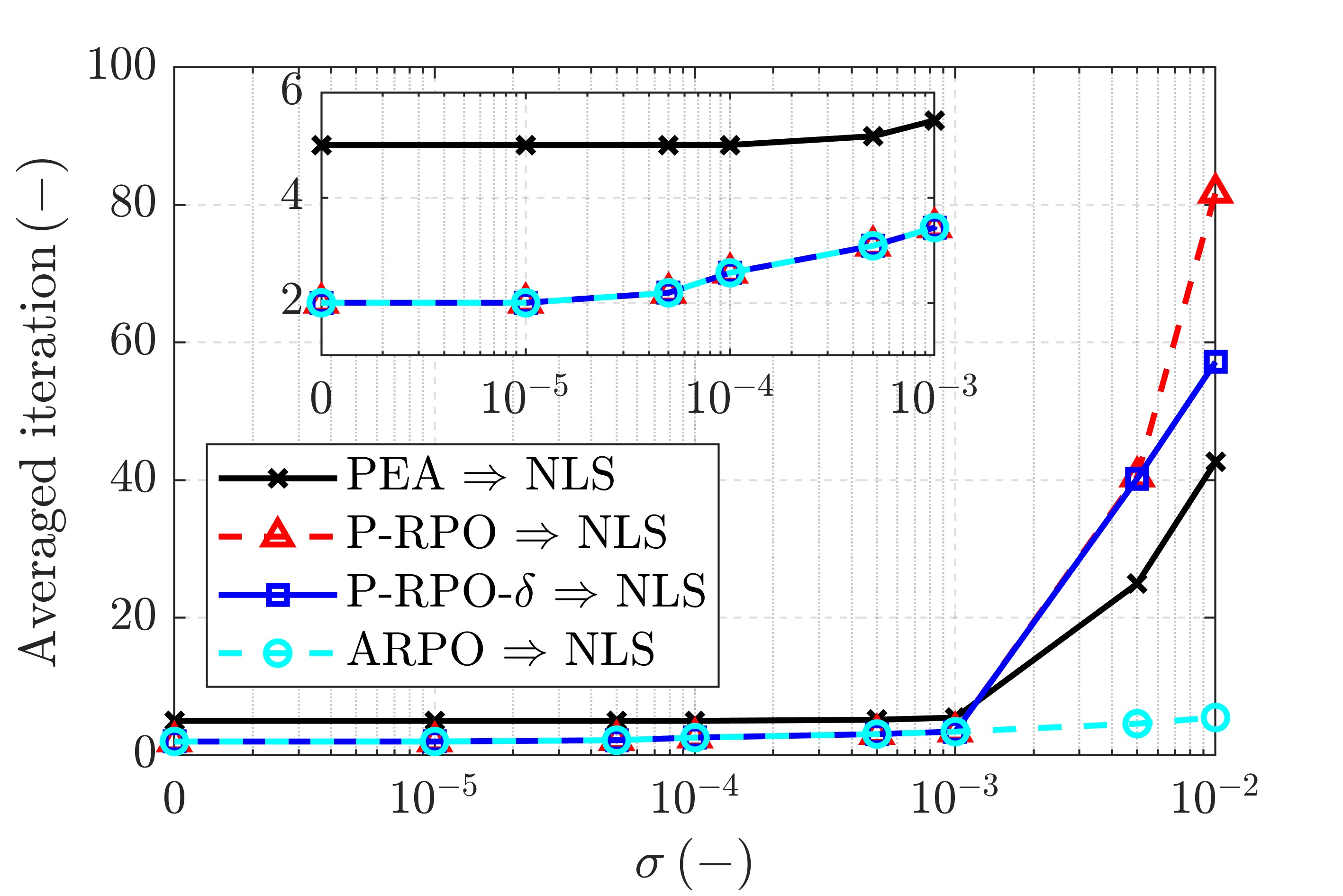}
    }
    \subfigure[]{
        \label{fig5d}
        \includegraphics[width=0.45\textwidth]{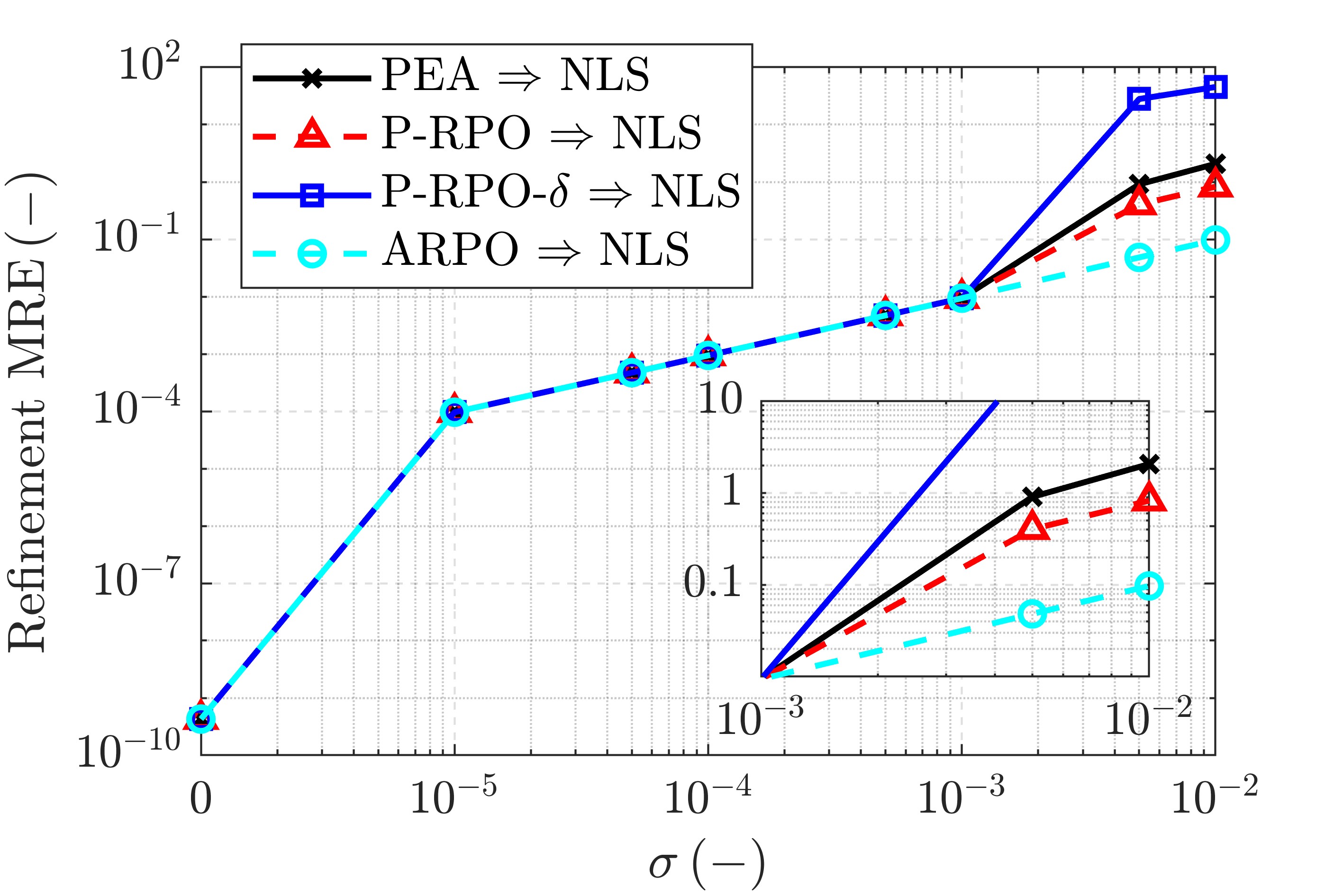}
    }
    \caption{Performance under different measurement noise levels (unweighted). \subref{fig5a} IROD MRE; \subref{fig5b} NLS convergence rate; \subref{fig5c} NLS averaged iterations; \subref{fig5d} NLS MRE.}
    \label{fig5}
\end{figure*}

This subsection first investigates the accuracy and convergence of different methods under different levels of measurement noise, characterized by the measurement error standard deviation $\sigma$, under the unweighted scheme. Three hundred MC runs are performed for each noise level. Only the first-order residual norm (\emph{i.e.}, $n$=1) is considered for the P-RPO, P-RPO-$\delta$ and ARPO, since the results in Sec.~\ref{sec4.2} show that it provides higher accuracy than the second-order form. Figure~\ref{fig5} summarizes the performance results under different measurement noise levels. As shown in Fig.~\ref{fig5a}, the P-RPO, P-RPO-$\delta$ and ARPO significantly outperform PEA in the IROD stage over the entire noise range. Among them, ARPO becomes the most accurate when the noise level is relatively high, especially for $\sigma>10^{-3}$.

Figures~\ref{fig5b}-\ref{fig5d} further show the performance of NLS refinement initialized by different IROD solutions under the unweighted scheme. For failed NLS runs, the iteration count in Fig.~\ref{fig5c} is set to 100. The improvement in IROD accuracy directly translates into better refinement performance, including higher convergence rates, fewer iterations, and lower refinement errors. In particular, ARPO achieves the highest convergence rate, the smallest averaged iteration number, and the lowest MRE after refinement. Similar trends are also observed for OLOD and are omitted for brevity.

Based on the above discussion, a particularly challenging case with $\sigma=10^{-2}$ is further considered. Note that $\sigma=10^{-2}$ corresponds to approximately $1.5^\circ$ of error at the $3\sigma$ level. Table~\ref{tab9} compares the performance of ARPO, L-RPO, and U-ARPO in this case. ARPO achieves the best IROD accuracy, reducing the MRE from 0.4489 and 0.4985 to 0.1876, i.e., by more than 50\% relative to the other two variants. This improvement also leads to markedly better downstream refinement performance. For example, in the NLS refinement, ARPO attains a 100\% convergence rate, whereas L-RPO converges in only 69\% of the trials, and the averaged number of refinement iterations is reduced from 34.78 to 5.49. These results show that both adaptive thresholding and constrained initialization play essential roles in achieving robust and accurate performance under severe noise conditions.

\begin{table*}[!h]
	\centering
	\caption{Performance of the ARPO, L-RPO, and U-ARPO in the challenging case (unweighted)}
	\label{tab9}
	\begin{tabular}{lccccccc}
        \hline\hline
        \multirow{3}{*}{Performance} & \multirow{2}{*}{IROD} & \multicolumn{6}{c}{Orbit refinement} \\ \cline{3-8} 
        & & \multicolumn{3}{c}{OLOD} & \multicolumn{3}{c}{NLS} \\ \cline{2-8} 
        & MRE & \multicolumn{1}{c}{MRE} & \multicolumn{1}{c}{Average iteration} & \multicolumn{1}{c}{CR} & \multicolumn{1}{c}{MRE} & \multicolumn{1}{c}{Average iteration} & CR \\ \hline
        ARPO & 0.1876 & \multicolumn{1}{c}{0.5650} & \multicolumn{1}{c}{49.0266} & \multicolumn{1}{c}{0.5833} & \multicolumn{1}{c}{0.0972} & \multicolumn{1}{c}{5.4900} & 1 \\
        L-RPO & 0.4489 & \multicolumn{1}{c}{0.6030} & \multicolumn{1}{c}{53.9800} & \multicolumn{1}{c}{0.5166} & \multicolumn{1}{c}{0.3781} & \multicolumn{1}{c}{34.7833} & 0.69 \\
        U-ARPO & 0.4985 & \multicolumn{1}{c}{0.5663} & \multicolumn{1}{c}{49.1666} & \multicolumn{1}{c}{0.5833} & \multicolumn{1}{c}{0.0972} & \multicolumn{1}{c}{5.9066} & 1 \\ \hline\hline
    \end{tabular}
\end{table*}

\begin{figure*}[!h]
    \centering
    \subfigure[]{
        \label{fig6a}
        \includegraphics[width=0.45\textwidth]{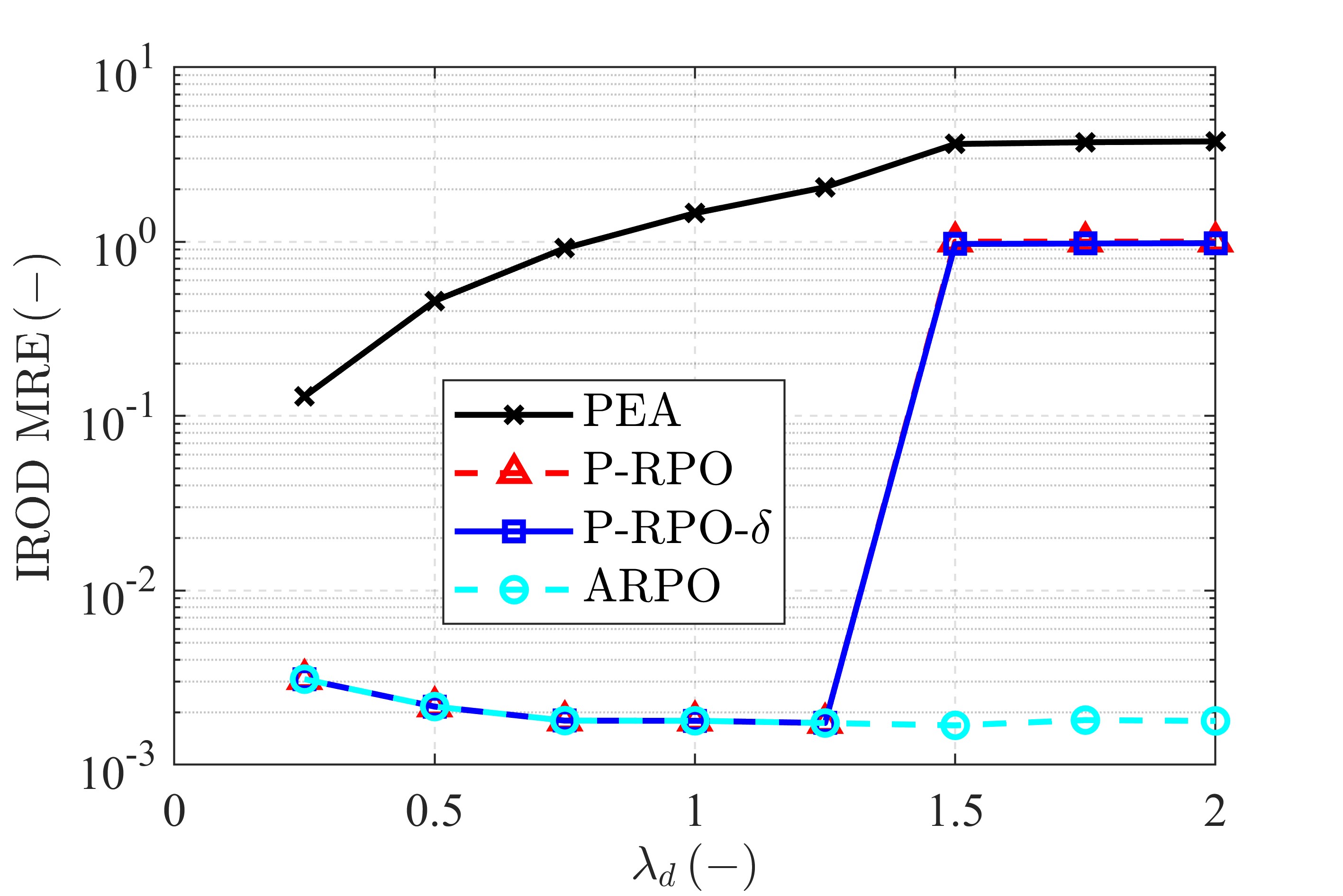}
    }
    \subfigure[]{
        \label{fig6b}
        \includegraphics[width=0.45\textwidth]{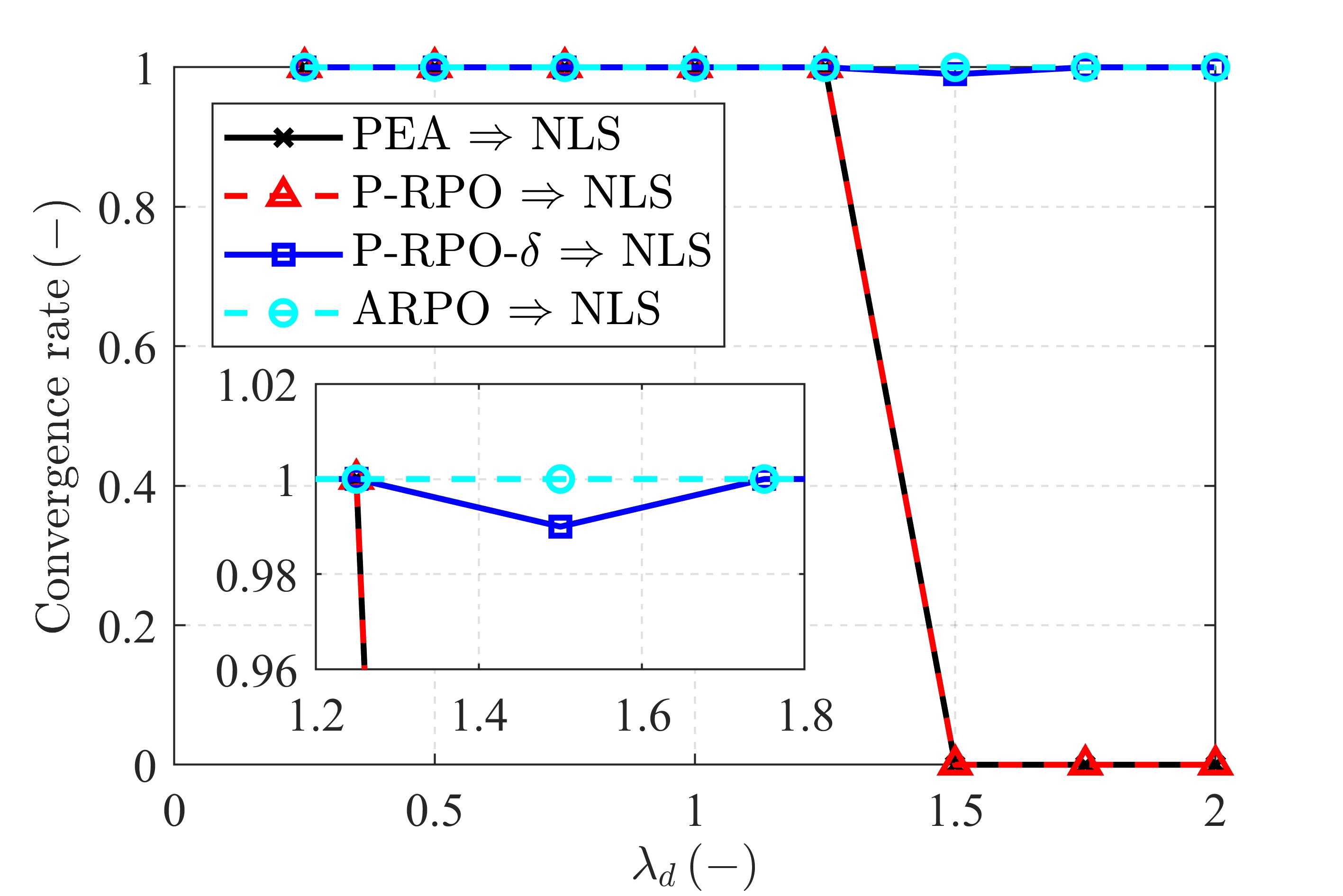}
    }
    \subfigure[]{
        \label{fig6c}
        \includegraphics[width=0.45\textwidth]{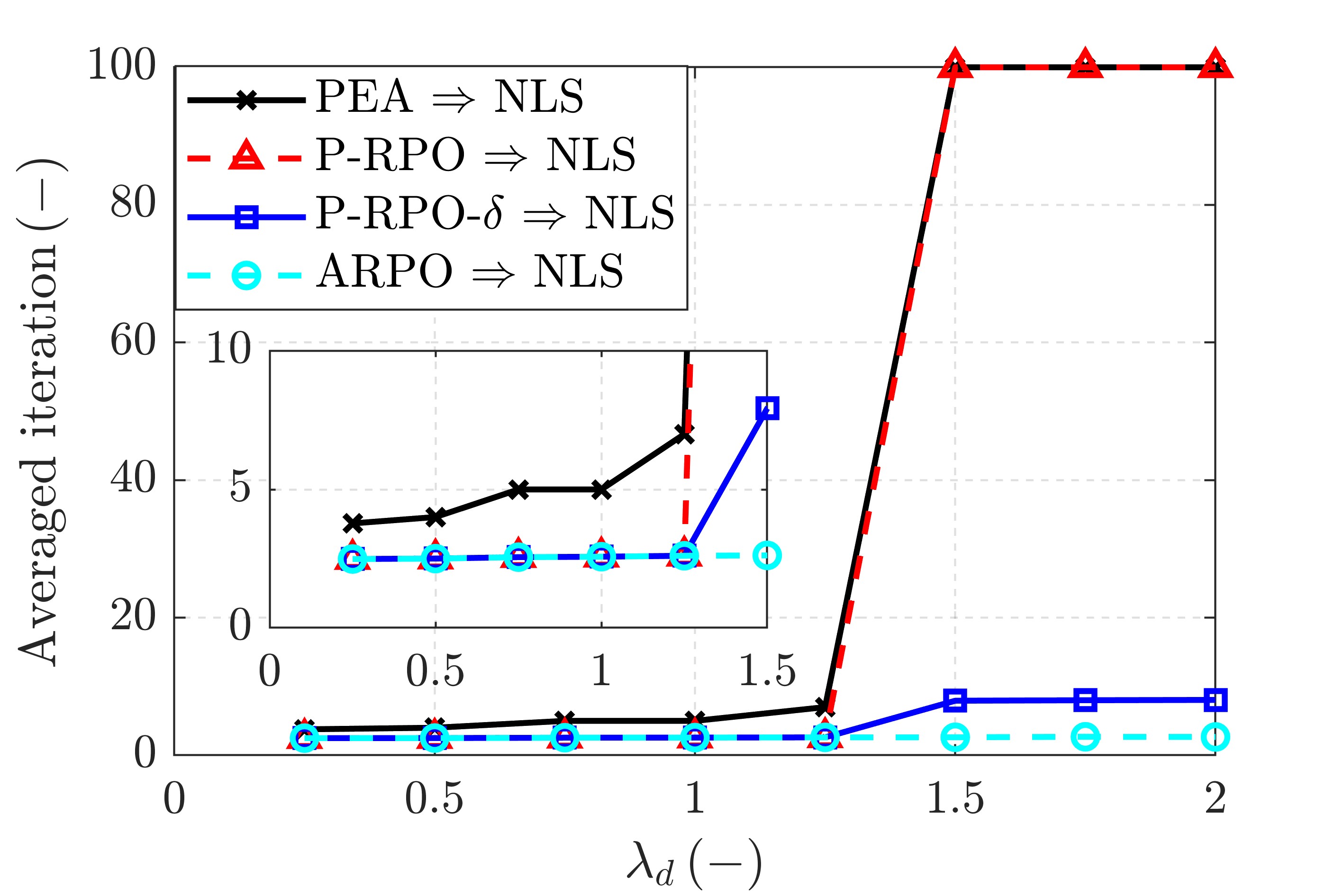}
    }
    \subfigure[]{
        \label{fig6d}
        \includegraphics[width=0.45\textwidth]{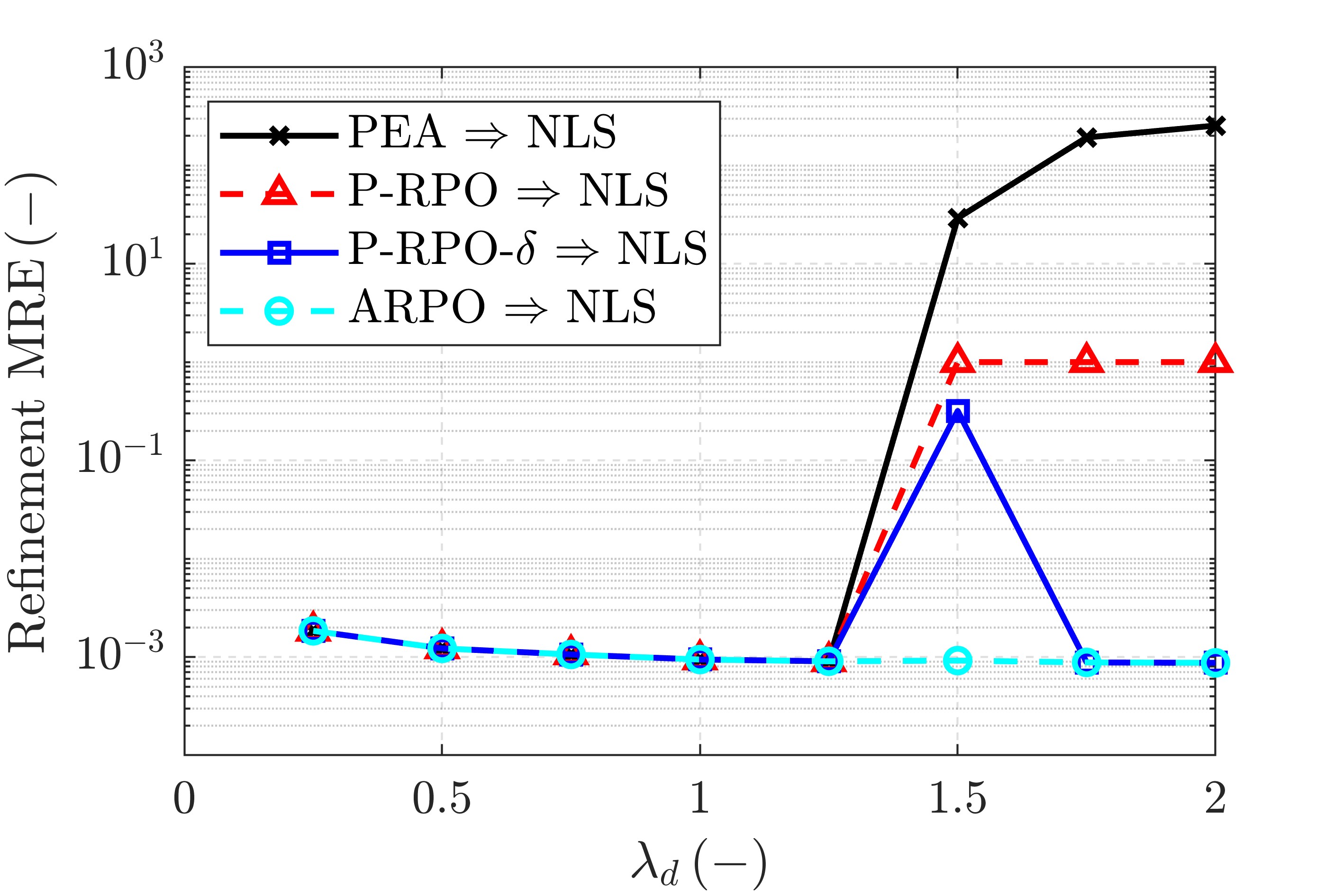}
    }
    \caption{Performance under different initial-distance scaling factors $\lambda_d$ (unweighted). \subref{fig6a} IROD MRE; \subref{fig6b} NLS convergence rate; \subref{fig6c} NLS averaged iterations; \subref{fig6d} NLS MRE.}
    \label{fig6}
\end{figure*}

The effect of the initial relative distance is further investigated under the unweighted scheme. A scaling factor $\lambda_d$ is introduced such that the initial relative state is replaced by $\delta \boldsymbol{x}_0 \gets \lambda_d \delta \boldsymbol{x}_0$. The value of $\lambda_d$ varies from 0.25 to 2.00 with an interval of 0.25. For each $\lambda_d$, 300 MC runs are performed with the measurement noise fixed at $\sigma=10^{-4}$. Figure~\ref{fig6} summarizes the corresponding robustness results. As shown in Fig.~\ref{fig6a}, the three RPO-based methods exhibit similar IROD accuracy when $\lambda_d \le 1.25$. However, when $\lambda_d \ge 1.25$, the non-adaptive methods P-RPO and P-RPO-$\delta$ can no longer reliably recover the desired solution, whereas ARPO remains accurate over the entire range of $\lambda_d$.

Figures~\ref{fig6b}-\ref{fig6d} further show the performance of NLS refinement initialized by the different IROD solutions. The degradation of the non-adaptive IROD methods at large $\lambda_d$ leads to lower convergence rates, more refinement iterations, and larger refinement errors. By contrast, ARPO maintains a 100\% convergence rate throughout the tested range and consistently achieves the smallest averaged iteration number and refinement error. Similar trends are also observed for OLOD and are omitted for brevity. These results indicate that the adaptive adjustment of the zero-avoidance threshold is essential for maintaining robustness as the initial relative distance increases.

\begin{figure}[!h]
	\centering
	\subfigure[]
	{
		\label{fig7a}
		\centering
		\includegraphics[width=0.45\textwidth]{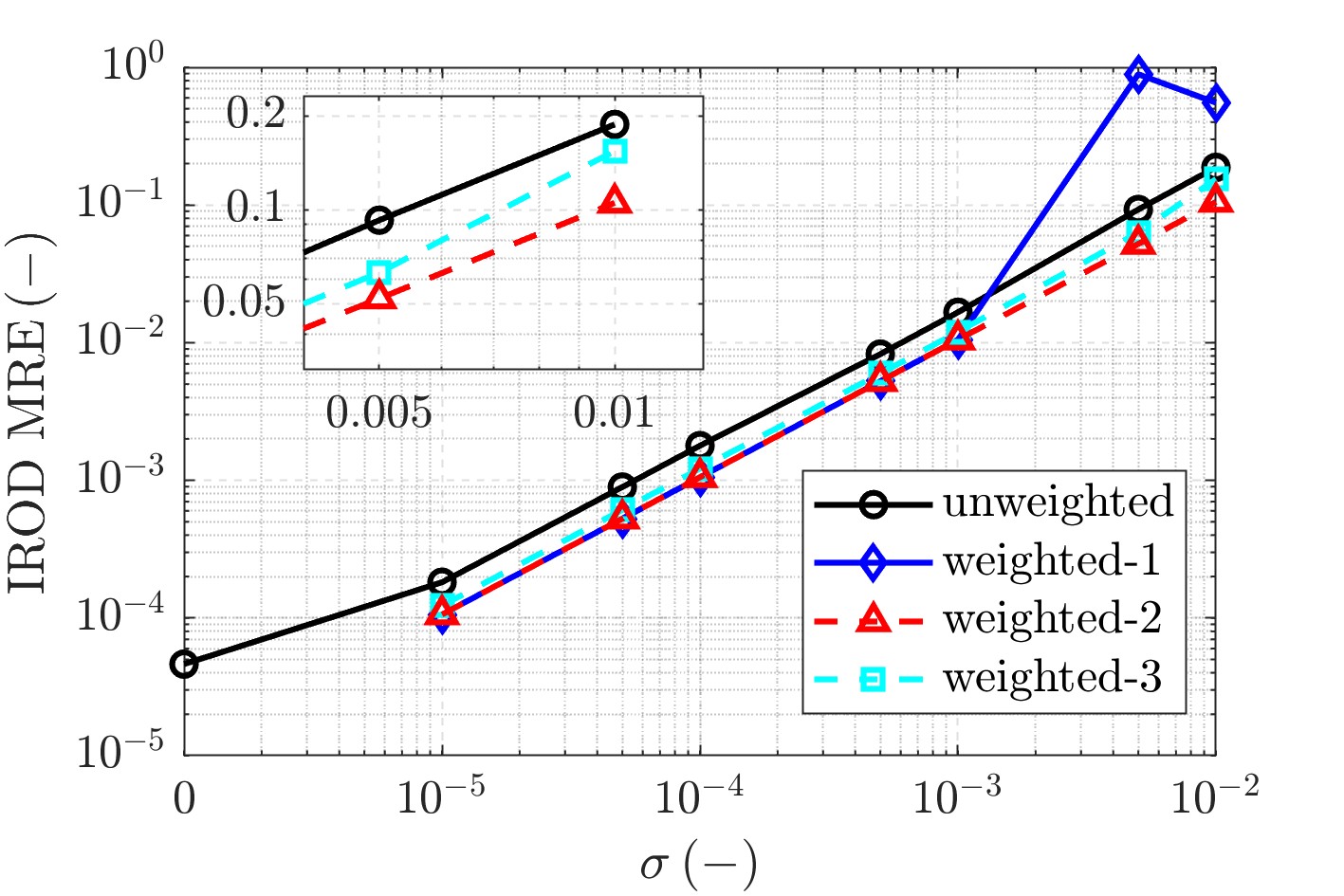}
	}
	\subfigure[]
	{
		\label{fig7b}
		\centering
		\includegraphics[width=0.45\textwidth]{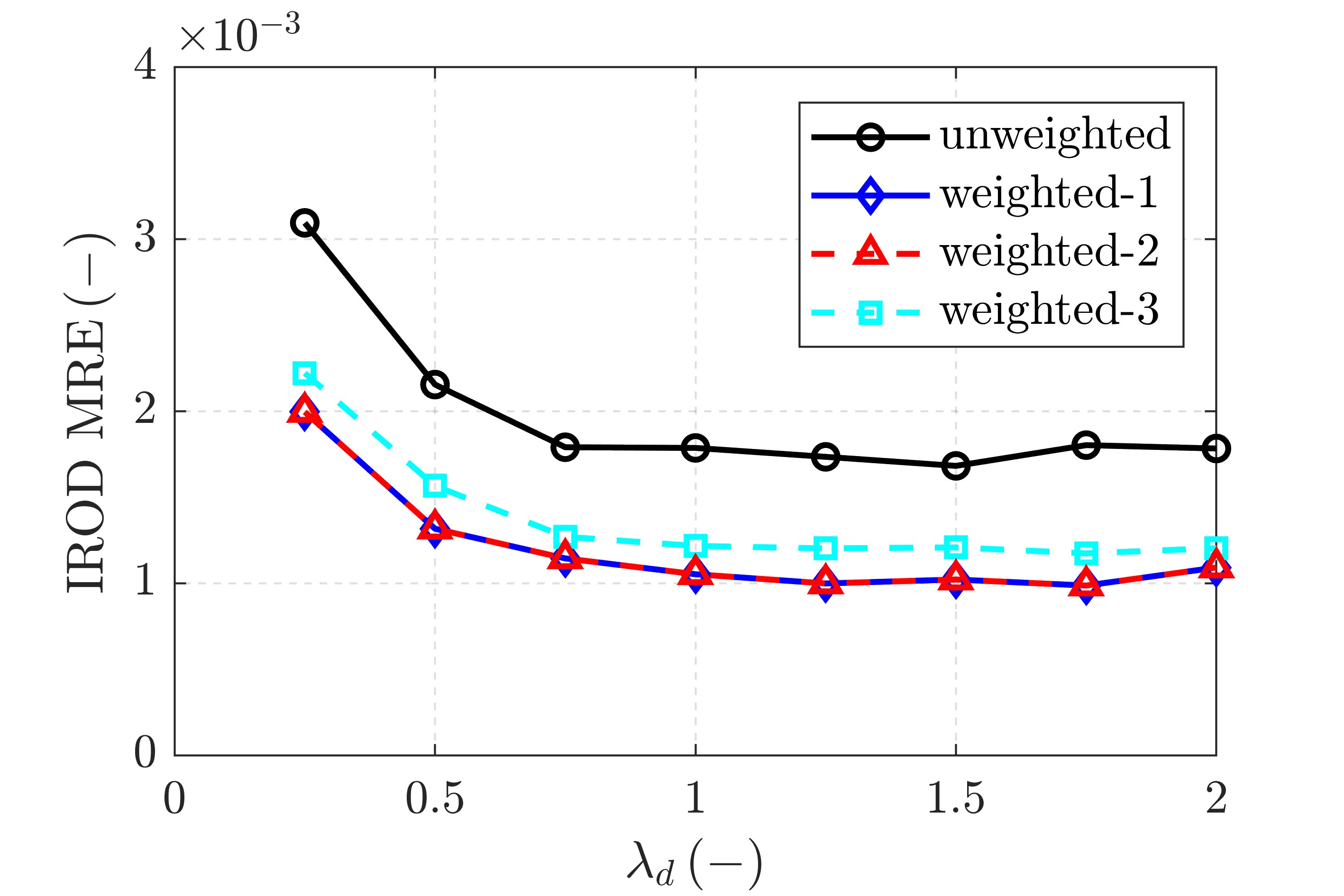}
	}
	\caption{Robustness analysis results of ARPO under weighted and unweighted schemes. \subref{fig7a} Effects of $\sigma$. \subref{fig7b} Effects of $\lambda_d$.}
	\label{fig7}
\end{figure}

Figure~\ref{fig7} compares the robustness of ARPO under the unweighted and weighted schemes. As shown in Fig.~\ref{fig7a}, all three weighted strategies outperform the unweighted one when $\sigma \le 10^{-3}$. For larger noise levels, however, the 3-D weighting strategy becomes markedly less stable, whereas the introduced reduced-order weighting and its simplified version remain stable and continue to outperform the unweighted scheme. Between the two introduced strategies, the reduced-order weighting is consistently more accurate, while the simplified one remains slightly less accurate due to the additional approximation (as discussed in Table~\ref{tab1}). This difference is most evident at $\sigma=10^{-2}$, where the IROD MREs of the unweighted, 3-D weighting, reduced-order weighting, and simplified reduced-order weighting schemes are 0.1876, 0.5527, 0.1053, and 0.1542, respectively. Relative to the other three schemes, the reduced-order weighting reduces the IROD MRE by about $43.9\%$, $81.0\%$, and $31.7\%$, respectively.

Figure~\ref{fig7b} further shows the effect of the initial-distance scaling factor $\lambda_d$ with the noise level fixed at $\sigma=10^{-4}$. In this case, all three weighted strategies outperform the unweighted one, while the 3-D and reduced-order weighting methods are nearly identical and both slightly outperform the simplified weighting. This indicates that, under low-noise conditions, varying $\lambda_d$ mainly challenges the zero-avoidance mechanism rather than the singularity handling of the weighting matrix; since ARPO can already adaptively select an appropriate threshold, the remaining differences among the weighting strategies are primarily due to their modeling accuracy.

\begin{figure}[!h]
	\centering
	\subfigure[]
	{
		\label{fig8a}
		\centering
		\includegraphics[width=0.45\textwidth]{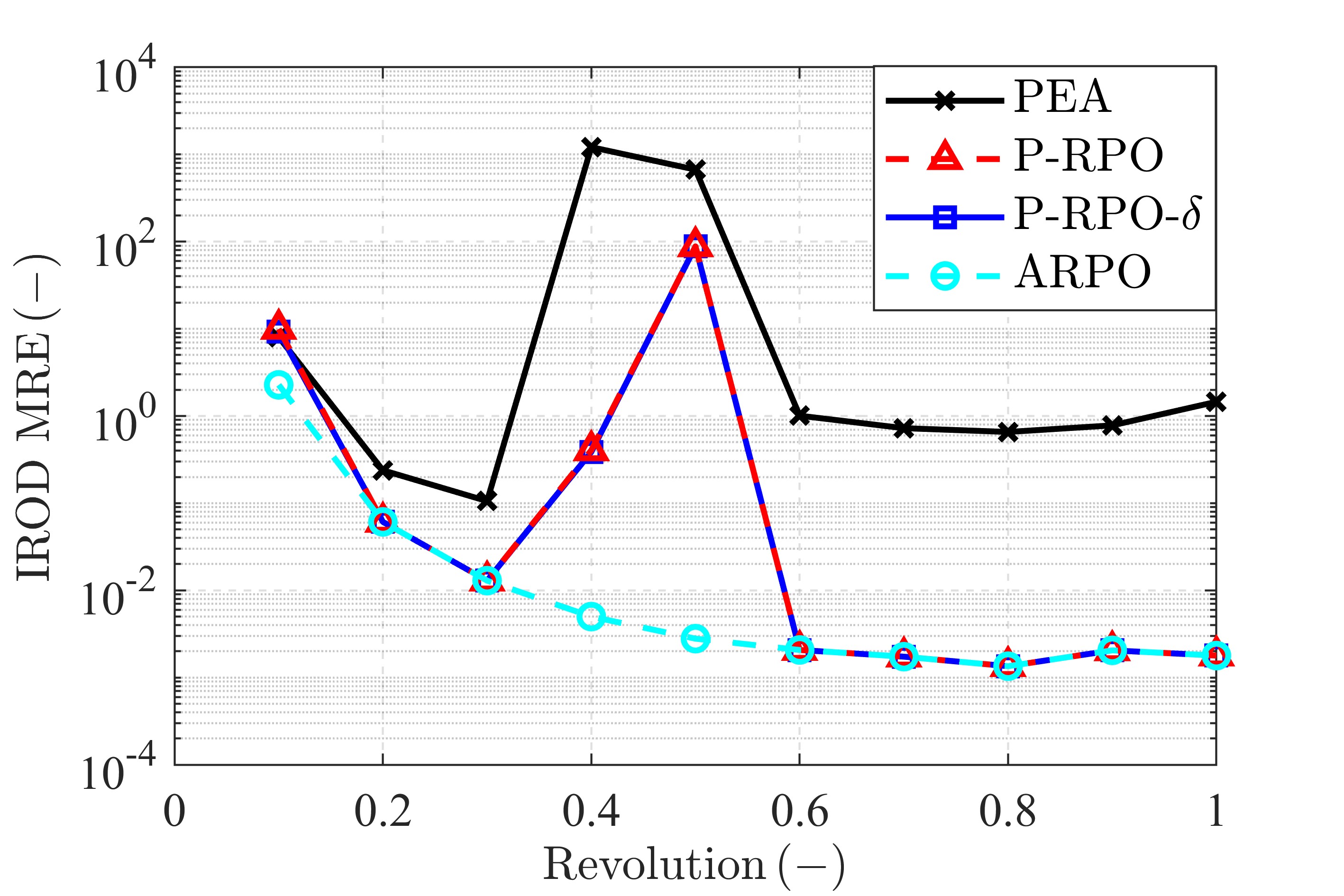}
	}
	\subfigure[]
	{
		\label{fig8b}
		\centering
		\includegraphics[width=0.45\textwidth]{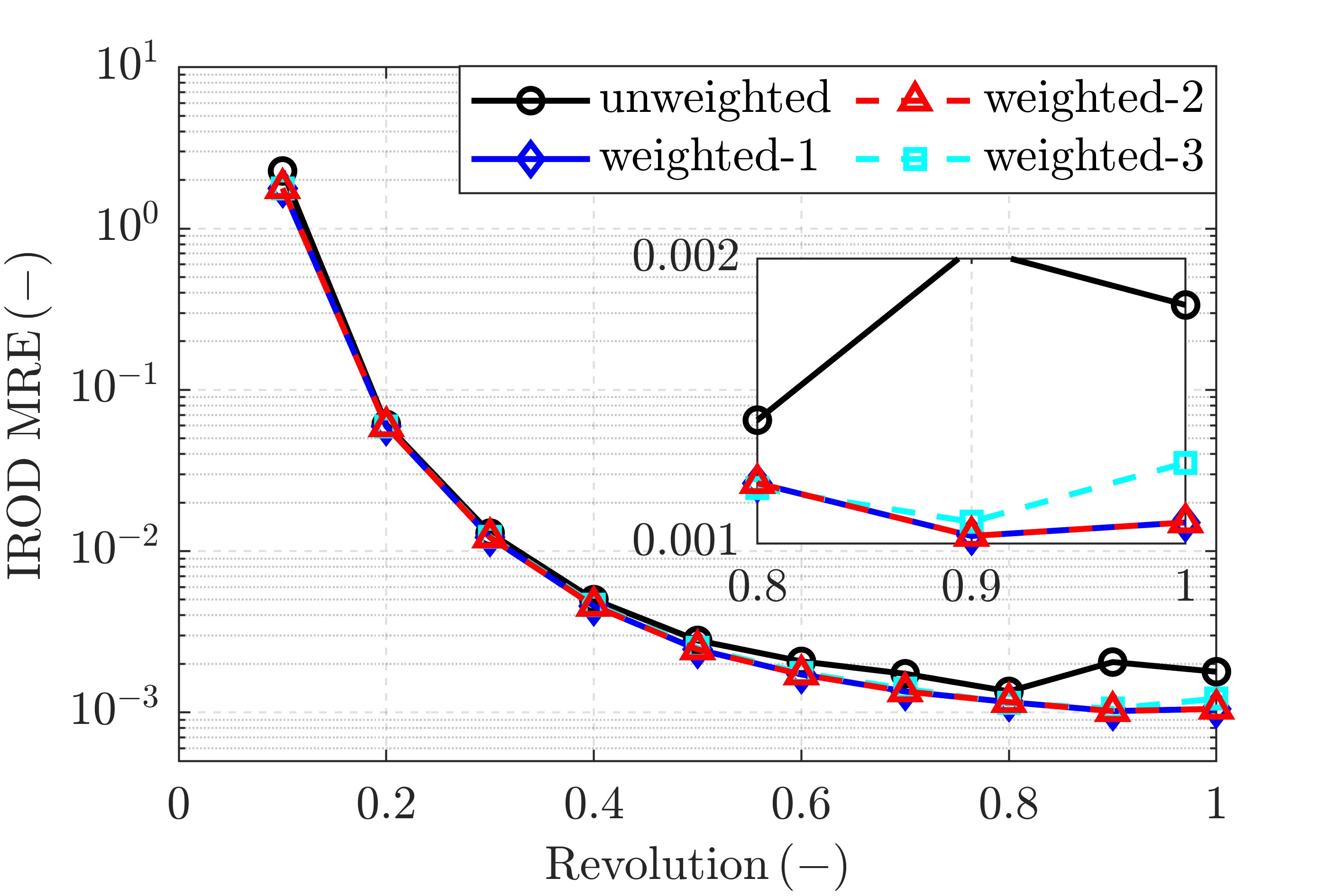}
	}
	\caption{Performance under different observation-arc lengths. \subref{fig8a} Unweighted methods. \subref{fig8b} Weighted and unweighted ARPOs.}
	\label{fig8}
\end{figure}

The effect of the observation-arc length is also investigated. 
The observation arc varies from 0.1$T$ to 1$T$, with 10 measurements and a fixed noise level of $\sigma = 10^{-4}$ in each case.
Three hundred MC runs are performed for each case. 
% As shown in Fig.~\ref{fig8a}, within the unweighted scheme, ARPO consistently outperforms PEA, P-RPO, and P-RPO-$\delta$, and remains stable over the entire range of observation-arc lengths. 
% In contrast, PEA exhibits pronounced error growth for some intermediate observation arcs (particularly around $0.5T$), which is consistent with the geometric sensitivity related to cross-track motion transfer noted in Ref.~\cite{Kulik2024}. Since P-RPO and P-RPO-$\delta$ use the PEA solution for initialization, their errors increase accordingly.
As shown in Fig.~\ref{fig8a}, within the unweighted scheme, ARPO consistently outperforms PEA, P-RPO, and P-RPO-$\delta$, and remains stable over the entire range of observation-arc lengths. Its advantage is particularly evident in two regimes. First, for very short observation arcs, ARPO provides a clear accuracy improvement. For example, at $0.1T$, the MRE of ARPO is about $2.27\%$, whereas the other three methods all have MREs of about $10\%$. Second, PEA exhibits pronounced error growth for some intermediate observation arcs (particularly around $0.5T$), which is consistent with the geometric sensitivity related to cross-track motion transfer noted in Ref.~\cite{Kulik2024}. Since P-RPO and P-RPO-$\delta$ use the PEA solution for initialization, their errors increase accordingly in these cases.
Figure~\ref{fig8b} further shows that, among the three weighting strategies, the reduced-order weighting and the 3-D weighting achieve nearly identical accuracy and both outperform the simplified and unweighted schemes. 
The reduced-order weighting does not show a clear advantage over the 3-D weighting in this comparison, because the measurement-noise level is not sufficiently severe for the singularity issue of the 3-D strategy to become dominant. Its main advantage is manifested under large-noise conditions, as shown in Fig.~\ref{fig7a}.

%%%%%%%%%%%%%%%%%%%%%%%%%%%%%%%%%%%%%%%%%%%%%%%%%%%%%%%%%%%%%%%%%%%%%%%%%%%%%%%%%%%%%
%%%%%%%%%%%%%%%%%%%%%%%%%%%%%%%%%%%%%%%%%%%%%%%%%%%%%%%%%%%%%%%%%%%%%%%%%%%%%%%%%%%%%
\section{Conclusion} \label{sec5}
Accurate angles-only initial relative orbit determination (IROD) solutions can be obtained by combining high-order Taylor polynomial approximation with cross-product-residual minimization, while the spurious zero solution can be effectively avoided through constrained optimization and adaptive thresholding.
Compared with the baseline methods, the introduced method improves the IROD accuracy by about three orders of magnitude and also reduces the downstream refinement cost by providing substantially more accurate initial estimates. 
In addition, it consistently delivers high-quality IROD solutions and enables nonlinear refinement to maintain a 100\% convergence rate in the robustness analysis, outperforming the baseline methods and the non-adaptive variants under severe noise, large initial distances, short observation arcs, and sensitive geometric configurations.
A reduced-order weighting strategy is further shown to improve estimation accuracy while structurally removing the singularity of the conventional 3-D weighting scheme. Compared with the unweighted scheme, the introduced weighting strategy improves the IROD accuracy by about 43\% in the nominal case. Under large-noise conditions, it remains stable and outperforms the conventional 3-D weighting by about 81\%.

\section*{Acknowledgments}
This work was supported by the National Natural Science Foundation of China (No. 62394353 and No. 124B2049) and the High-level Talent Support Program (No. T2023191). Xingyu Zhou also thanks Zeno Pavanello for his helpful discussion on recursive polynomial optimization.

\bibliography{sample}

\end{document}